\documentclass[%
aps,
prd, 
twocolumn,
superscriptaddress,
nofootinbib,
amsmath,amssymb,amsfonts,
]{revtex4-2}

\usepackage{aas_macros}
\usepackage[T1]{fontenc}
\usepackage[english]{babel}
\usepackage{bm} 
\usepackage{empheq}
\usepackage{xcolor}
\usepackage[breaklinks=true,colorlinks,citecolor=blue,linkcolor=blue,urlcolor=blue]{hyperref}

\usepackage{enumitem} 

\usepackage{amsmath,amsthm,amssymb} 
\usepackage{mathtools,bbm}

\newcommand{\reals}{\mathbb{R}} 
\newcommand{\ee}{\mathrm{e}} 

\renewcommand{\vec}[1]{\boldsymbol{#1}} 
\newcommand{\mat}[1]{\mathbf{#1}} 

\newcommand{\idmat}[1]{\mathbbm{1}_{#1}} 
\DeclareMathOperator{\diag}{diag} 
\DeclareMathOperator{\antidiag}{antidiag} 

\newcommand{\tr}[1]{\prescript{t}{}{#1}} 

\newcommand{\act}{\mathcal{S}} 
\newcommand{\lagstd}{\mathcal{L}} 
\newcommand{\hamstd}{\mathcal{H}} 
\newcommand{\lagext}{\Lambda} 
\newcommand{\hamext}{\mathcal{A}} 

\newcommand{\ts}{t_s} 
\newcommand{\te}{t_e} 

\newcommand{\q}{q}
\newcommand{\p}{p}
\newcommand{\qdot}{\dot{\q}}
\newcommand{\vq}{\vec \q}
\newcommand{\vqdot}{\dot{\vq}}
\newcommand{\vqddot}{\ddot{\vq}}
\newcommand{\vp}{\vec \p}

\newcommand{\labelA}{{\uparrow}} 
\newcommand{\labelB}{{\downarrow}} 
\newcommand{\pathA}[2][]{{#2}_{\labelA #1}} 
\newcommand{\pathB}[2][]{{#2}_{\labelB #1}} 
\newcommand{\pathAB}[1]{{#1}_{\uparrow\joinrel\downarrow}} 
\newcommand{\pathinterchange}{(\uparrow\joinrel\downarrow)}
\newcommand{\pathany}[1]{{#1}_{\{\raisebox{-.1em}{\argdot}\}}} 
\newcommand{\pathmetric}[1][ab]{\eta_{#1}} 

\newcommand{\PL}{\mathrm{PL}}
\newcommand{\vorder}[1]{\mathcal{O}(-^{#1})}

\newcommand{\inner}[2]{\left\langle #1, #2 \right\rangle} 
\newcommand*{\argdot}{\makebox[1ex]{\textbf{$\cdot$}}} 
\DeclarePairedDelimiter\norm{\lVert}{\rVert}%

\newcommand{\dd}{\mathrm{d}} 
\newcommand{\dndot}[2]{\overset{\scriptscriptstyle (#2)}{#1}} 
\newcommand{\grad}{\vec{\nabla}} 
\newcommand{\laplac}{\nabla^2} 
\newcommand{\dext}[2]{\frac{\partial #1}{\partial #2\!\!}\,} 
\newcommand{\Dt}{\mathrm{D}_t}
\newcommand{\lPB}{\{\!\{}
\newcommand{\rPB}{\}\!\}}

\let\originalleft\left
\let\originalright\right
\renewcommand{\left}{\mathopen{}\mathclose\bgroup\originalleft}
\renewcommand{\right}{\aftergroup\egroup\originalright}

\usepackage[textsize=small]{todonotes}

\begin{document}

\preprint{APS/123-QED}

\title{Hamiltonian treatment of non-conservative systems}

\author{C.~Aykroyd}\email{christopher.aykroyd@obspm.fr}
\affiliation{LTE, Observatoire de Paris, Universit\'e PSL, CNRS, Sorbonne Universit\'e, LNE, 61 avenue de l'Observatoire, 75014 Paris, France}

\author{A.~Bourgoin}
\affiliation{LTE, Observatoire de Paris, Universit\'e PSL, CNRS, Sorbonne Universit\'e, LNE, 61 avenue de l'Observatoire, 75014 Paris, France}

\author{C. Le Poncin-Lafitte}
\affiliation{LTE, Observatoire de Paris, Universit\'e PSL, CNRS, Sorbonne Universit\'e, LNE, 61 avenue de l'Observatoire, 75014 Paris, France}

\date{\today}

\begin{abstract}
    We present a novel extension of Hamiltonian mechanics to nonconservative systems built upon the Schwinger-Keldysh-Galley double-variable action principle. Departing from Galley's initial-value action, we clarify important subtleties regarding boundary conditions, the emergence of the physical-limit trajectory, and the decomposition of the Lagrangian into conservative and dissipative sectors. 
    Importantly, we demonstrate that the redundant doubled configuration space admits a gauge freedom at the level of the canonical momenta that leaves the physical dynamics unchanged.
    From a Legendre transform, we construct the corresponding family of gauge-related nonconservative Hamiltonians; we show that virtually any classical initial-value problem can be embedded on our enlarged symplectic manifold, supplying the associated Hamiltonian and Lagrangian functions explicitly.
    As a further contribution, we derive a completely equivalent linear ``Lie'' formulation of the double-variable action and Hamiltonian which streamlines computations and renders transparent many structural properties of the formalism.
    
\end{abstract}

\maketitle


\section{Introduction}

Action principles stand as a foundational pillar in physics; they offer a compact, unifying viewpoint from which symmetries and conservation laws are naturally linked. Yet despite their prevalence in physics, they traditionally fail to provide adequate treatment of non-conservative systems, those involving forces that cannot be derived from a generalized potential. Many physical processes involve loss of energy (dissipation), stochasticity, or coupling to an external environment. Such processes may also be \emph{irreversible}, indicating that energy is never fully recovered once lost.\footnote{Dissipative processes are both \emph{non-conservative} and \emph{irreversible}; energy is permanently lost into the environment, accompanied by a total entropy increase. Meanwhile, non-conservative processes can also describe, for instance, open systems where information will eventually be recovered.} Often, non-conservative behaviour must be incorporated directly at the level of the equations of motion, restricting the advantages of any action-based approach. While one can often incorporate non-conservative processes directly into the \emph{equations of motion}, it is far more challenging to derive these same equations from an \emph{action principle}. Indeed, a widely held perspective is that dissipation and other forms of non-conservative dynamics inherently defy the neat structure of traditional Lagrangian or Hamiltonian mechanics. On the contrary, as we shall see, certain modifications to the action principle can allow non-conservative processes to be treated satisfactorily.



We can gain insight from Noether’s theorem, which establishes a direct connection between symmetries of the action and conserved quantities in a physical system.  Specifically, if the Lagrangian does not explicitly depend on time---it is invariant under continuous time translations---then conservation of Noether energy is guaranteed. Establishing an action principle which violates energy conservation therefore requires one of three routes:
\begin{enumerate}
\item A description of the system via a non-autonomous Lagrangian;
\item A dissociation between the physical and Noether energies; for example, in open systems exchanging energy with the environment, the conserved quantity might represent a combination of the physical energy of the system and the environment;
\item A modification of the structure of the action (e.g.\ introducing non-locality) or of the boundary conditions, so that the usual assumptions of the theorem are infringed.
\end{enumerate}
As we shall see in the discussion that follows, although the first item (1) provides an obvious pathway for incorporating the physics of non-conservative processes, it is not always desirable or feasible. In many scenarios, a combination of (2) and (3) proves far more flexible or general.

Another important theoretical result in this context was historically demonstrated by \citeauthor{Bauer1931}, and asserts that ``the equations of motion of a dissipative linear dynamical system with constant coefficients are not given by a variational principle’' \cite{Bauer1931}. While Bauer's statement seems discouraging, it, too, implicitly hinges on assumptions, which are imposed from the construction of the variational principle. Namely, the corollary holds when the action is a single scalar functional of the form
\begin{equation}\label{eq:conservative_action} 
    \act [ \vq ] = \int_{\ts}^{\te} \! \lagstd \left(\vq, \vqdot, \vqddot, \ldots, \dndot{\vq}{M} \right) \, \dd t \text,
\end{equation}
depending solely on the system’s variables and their integer-order derivatives $\dndot{\vec q}{\ell}, \ell \leq M$
, with the action variations vanishing at the endpoints. Conveniently, relaxing each of these conditions hints to a possible pathway to define an action principle compatible with non-conservative systems. Following this line of thought, several strategies have been proposed, which range from modifying the functional form of the action \cite{Gurtin1963,Gurtin1964,Tonti1973,Dargush2012,Dargush2012a}, to introducing additional degrees of freedom \cite{Bateman1931, Schwinger1961, Keldysh1964, Galley2013}, to incorporating fractional-order derivative dependencies into the Lagrangian \cite{Riewe1996, Riewe1997, Dargush2012, Dargush2012a}. In the following paragraphs, we will examine these approaches in detail, highlighting the primary challenges and the innovative solutions developed to overcome them.

\subsection{Early approaches to non-conservative action principles}

We begin with some of the most traditional strategies for restoring a variational structure in the presence of non-conservative forces.

\subsubsection{Explicit time dependency}

A straightforward approach is to introduce explicit time dependence into the Lagrangian, thereby creating systems whose Noether energy is not conserved. A classic example is the damped harmonic oscillator described by the Lagrangian
\begin{equation}
    \lagstd ( \q, \qdot, t ) = \frac{1}{2} \left( m \qdot^2 - k \q^2 \right) \ee^{\gamma t / m} \text, \label{eq:damped_harmonic_oscillator_action}
\end{equation}
which produces the correct equations of motion, but at the cost of obscuring the physical interpretation of canonical variables and conserved quantities \cite{Riewe1996}. The corresponding time-dependent Hamiltonian is known as the Caldirola-Kanai Hamiltonian \cite{Caldirola1941, Kanai1948}.
Another practical limitation of this method is its strong dependence on the system being described. Constructing a compatible Lagrangian often involves ``reverse-engineering'' the Lagrangian---by ensuring the resulting Euler–Lagrange equations match the known dynamics---and can require an ``integrating factor'' or ad-hoc algebraic manipulations\footnote{The so-called \emph{Helmholtz conditions} establish the requirements for when a system of equations can be recognised \emph{as is} as the Euler–Lagrange equations of some Lagrangian. They do not, however, account for nontrivial transformations—such as multiplicative factors, mixing of equations, or changes of variables—that might recast the system into a variational form.}. In general, there is no universal guarantee that such a multiplier can even be found for arbitrary dissipative systems. 
Furthermore, once the Lagrangian explicitly depends on time, the relationship between $\vp$ and $\vqdot$ may become non singular and the Legendre transform to a Hamiltonian formulation, $\lagstd(\vq, \vqdot, t)$ to $\hamstd(\vq, \vp, t)$, ill-defined.

Another related strategy to the ad-hoc prescription is to model the environment explicitly in the action; then, by integrating out environmental degrees of freedom one derives an effective open system. 
For instance, modeling the environment as a collection of harmonic oscillators and applying a statistical limit can lead to Langevin- or Fokker-Plank-type equations (see e.g., \citet{Calzetta2008}). In the context of General Relativity (GR), a similar procedure---integrating out metric perturbations---produces dissipative post-Newtonian equations of motion (e.g.\ the 2.5PN radiation-reaction terms), even though the complete underlying system is itself Lagrangian (however, it should be noted that defining a conserved ``canonical energy'' in GR can be subtle). Unfortunately, this type of technique cannot always be directly applied to the action, if the form of the action inherently disallows dissipative effects. The reasons for this inability will be further explored in Sec.\ \ref{sec:intro:bvp}. In general, one must instead be content with a dissipative description of the equations of motion.
In a nutshell, while these methods are effective in certain contexts, they are most suited to scenarios where environmental degrees of freedom can be explicitly solved.

\subsubsection{Rayleigh dissipation function}

In contrast to methods that rely on explicit time dependence or environmental modeling, strategies that treat dissipation dynamically and are broadly applicable to non-conservative systems are often more desirable. One of the earliest and most influential approaches in this direction is the \emph{Rayleigh dissipation function} \cite{Rayleigh1896}. Originally introduced as a quadratic form of velocities, $\mathcal F(\vq, \vqdot) = \gamma \qdot^2$, the Rayleigh function provides a way to incorporate damping forces into the dynamics. Specifically, the gradient of $\lagstd$ with respect to the velocities yields the dissipative forces, which are then added to the Euler-Lagrange equations:
\begin{equation}
\frac{\dd}{\dd t} \frac{\partial \lagstd}{\partial \vqdot} -  \frac{\partial \lagstd}{\partial \vq} +  \frac{\partial \mathcal F}{\partial \vqdot} = 0 \text. \label{eq:Rayleigh_diss_fun}
\end{equation}
While extensions of the Rayleigh function have been proposed to accommodate more complex dissipation mechanisms \cite[see][]{Virga2015}, its utility remains constrained.
Specifically, the Rayleigh dissipation function is not integrated into the action itself; instead, its contribution is introduced ad hoc into the equations of motion. Although it has been shown that Eq.\ \eqref{eq:Rayleigh_diss_fun} can be derived from a \emph{variational} principle, it does not arise from a single scalar action functional \cite{Dargush2012}, which undermines the generality of the approach.

\subsection{Classical action principle as a boundary-value problem}\label{sec:intro:bvp}

It turns out that a major hindrance in extending action principles to non-conservative systems arises from the fact that these principles are traditionally framed as boundary-value problems. As we shall see, when the underlying physics requires an initial-value treatment---as is the case for non-conservative systems---this hindrance turns irreconcilable. As highlighted by multiple works \cite{Gurtin1963, Tonti1973, Riewe1996, Dargush2012, Galley2013}, resolving this incompatibility is ultimately a key to providing a general action principle valid for non-conservative systems.

The classical action principle states that the physical path $\vec q(t)$ taken by a Lagrangian system between two specified states $\vec q(\ts) = \vec q_s$ and $\vec q(\te) = \vec q_e$ at times $\ts$ and $\te$ is the one for which the action functional $\act$ is stationary:
\begin{equation}\label{eq:conservative_action_principle}
\begin{cases}
    \delta \act[\vq, \delta \vq] = 0 \text, \quad \forall \, \delta \vq \text, \\
    \vq(\ts) = \vq_s \text, \\
    \vq(\te) = \vq_e \text,
\end{cases}
\end{equation}
where henceforth we restrict the Lagrangian to depend on up to first-order derivatives.
Such formulation describes a \emph{boundary value problem} (BVP), since both initial and final states are fixed. As highlighted by \cite{Tonti1973, Dargush2012} and others, this action principle generally leads to the \emph{equivalent} Euler-Lagrange equations which inherit these boundary conditions:
\begin{equation}\label{conservative_EL_bvp}
\begin{cases}\displaystyle
    \frac{\partial \lagstd}{\partial \vq} - \frac{\dd}{\dd t} \frac{\partial \lagstd}{\partial \dot{\vec q}} = 0 \text, \\
    \vec q(\ts) = \vec q_s \text, \\
    \vec q(\te) = \vec q_e \text.
\end{cases}
\end{equation}
However, in physics, we often require a description of an \emph{initial value problem} (IVP), where the system's state is specified at an initial time, and its future evolution is predicted. The traditional action principle \eqref{eq:conservative_action_principle} conflicts with this scenario because it necessitates conditions at both initial and final times.  The typical workaround is to derive the equations of motion from the BVP principle \eqref{eq:conservative_action_principle} and then artificially impose initial value conditions on \eqref{conservative_EL_bvp}, \emph{even though boundary conditions were used to derive these equations}.
As demonstrated by \citeauthor{Tonti1973} \cite{Tonti1973}, the artificial conversion of a BVP into an IVP is valid only when the Euler operator is self-adjoint (more details below), but fails more generally, particularly for nonconservative systems. 

Another way to understand the inherent conflict between BVPs and IVPs is through Green’s functions. In the linear (and by extension, in the perturbed, quasi-linear) case, boundary value problems are solved using \emph{symmetric} Green’s functions---the impulse response that satisfies boundary conditions at both initial and final times. This symmetry enforces time-reversible dynamics. In contrast, initial-value problems are described by \emph{retarded} (causal) Green’s functions, which encode a clear direction of time and can accommodate nonconservative effects. Using a symmetric Green’s function in a scenario where the system is supposed to evolve forward from an initial state means that the future is inappropriately constrained by boundary conditions imposed at the final time.\footnote{This distinction between symmetric and retarded Green’s functions also has profound implications for the construction of \emph{effective actions}, which are often derived by integrating out a subsystem (e.g., environmental variables) from a larger system. In this paradigm, the use of the symmetric Green’s functions is critical in order to satisfy the BVP action formulation. Unfortunately, the resulting effective action will inherently suppress dissipative effects, since the BVP and symmetric Green’s function enforce time-reversibility \cite[see][]{Galley2013}. In contrast, the retarded Green’s function, which is appropriate for IVPs, cannot be incorporated into a traditional action principle \cite[see e.g.\ the appendix in][]{Damour2016}. This explains why effective actions generally fail to capture dissipative behavior.}
Non-conservative dynamics are characterised by causal and irreversible behaviour; the system's state at a future time is inherently path-dependent. Such attribute underlies why nonconservative phenomena cannot be framed as BVPs.

To address these limitations within the action framework, 
it is necessary to reformulate the principle to be directly compatible with IVPs rather than BVPs. Such a reformulation not only resolves this shortcoming of classical physics but also naturally enables the treatment of nonconservative systems. A variety of strategies have already been proposed in this regard.

\subsubsection{Convolved action principles}

A promising option for nonconservative \emph{linear} variational principle involves redefining the action using convolution operators. This strategy was pioneered by \citeauthor{Gurtin1963} \cite{Gurtin1963, Gurtin1964}, who proposed transforming linear IVPs into equivalent integro-differential BVPs whose equations were expressed using convolution integrals. His method ensured that initial conditions were inherently satisfied due to the structure of the equations, and that the equations of motion could be derived from a traditional BVP action framework. \citeauthor{Gurtin1963} demonstrated the effectiveness of his technique by applying it to the heat and wave equations. Later, \citeauthor{Tonti1973} \cite{Tonti1973} advanced this idea with a distinct convolution formulation that enforced symmetry in the resulting differential operator.

Central to these approaches is the variational structure of action principles. For any system, the action principle will lead to a variational form of the kind
\begin{equation}
    \delta \act[\vec q, \delta \vec q] = \langle E[\vec q], \delta \vec q \rangle = 0 \text,  \quad \forall \delta \vec q \text, \label{eq:EL-bilinear_form}
\end{equation}
where $\inner{\argdot}{\argdot}$ is a bilinear form, and $E$ is the differential operator---often called the Euler operator---from which the Euler-Lagrange equations $E[\vec q] = 0$ are derived. The bilinear form arises after integration by parts and is implicitly determined by the structure of the action functional. Namely, in traditional mechanics with the action \eqref{eq:conservative_action}, the bilinear form corresponds to the standard integral inner product
\begin{equation}
    \inner{\vec u}{\vec v}_{\mathrm s} = \int_0^T \vec u(t) \cdot \vec v(t) \, \dd t \text.
\end{equation}
This bilinear form is well-suited for describing BVPs and conservative systems because, when applying integration by parts, the integration boundaries vanish, so that
\begin{equation}
    \inner{ \dot{\vec u } }{ \vec v }_{\mathrm s} = - \inner{ \vec u }{ \dot{\vec v} }_{\mathrm s} \text. 
\end{equation}
In other words, time-derivative terms $\dd/\dd t$ in the Euler operator are \emph{anti-symmetric} with respect to the standard inner product, whereas the square $\dd^2/\dd t^2$ is \emph{symmetric}. Symmetric (or self-adjoint) linear operators are particularly important in variational principles because they guarantee the existence of an action leading to $E[\vec q] = 0$. More precisely, when the Euler operator is non-linear, if its functional derivative is symmetric, then the system can be derived from a Lagrangian \cite{Tonti1973}.%
\footnote{However, it's important to note that while a symmetric operator indicates the existence of a Lagrangian, the converse is not always straightforward. In some cases, equations of motion may require manipulation to reveal an underlying symmetric operator.
Consider, for example, the damped harmonic oscillator governed by:
\begin{equation}
    m \ddot q(t) + \gamma \dot{q}(t) + k q(t) = 0 \text. \label{eq:damped_oscillator} 
\end{equation}
The naive choice of Euler operator $E = m\, \partial_t^2 + \gamma\, \partial_t + k$ is not symmetric due to the presence of the first-derivative term (which is anti-symmetric). However, by multiplying Eq.\ \eqref{eq:damped_oscillator} by an integrating factor $e^{\gamma t / m}$, the new operator $E' = e^{\gamma t / m} E$ becomes symmetric:
\begin{equation} 
    \langle E'(q), \delta q \rangle = \int_0^T e^{\gamma t / m} \left( m \dot{q} \delta \dot{q} + k q \delta q \right) \dd t = \langle q, E'(\delta q) \rangle \text,
\end{equation}
where we have applied integration by parts. This manipulation allows the variation to be extracted from the integral, from whence we may derive an action functional:
\begin{align}
    \delta \act[q, \delta q] &= \frac{1}{2} \left( \langle E'(q), \delta q \rangle + \langle E'(\delta q), q \rangle \right) \\
    &= \delta \int_0^T \frac{1}{2} e^{\gamma t / m} \left( m \dot{q}^2 + k q^2 \right) \dd t \text. \nonumber
\end{align} 
In the above, we recognise the Lagrangian for the damped harmonic oscillator \eqref{eq:damped_harmonic_oscillator_action}. This manipulation demonstrates how ad-hoc adjustments can enforce symmetry for specific systems.}

When dealing with IVPs, however, the standard inner product fails: boundary terms do not vanish since only initial conditions are specified. To address this, \citeauthor{Tonti1973} proposed replacing the standard inner product with a convolution:
\begin{equation}
    \inner{\vec u}{\vec v}_{\mathrm c} = \int_0^T \vec u(t) \cdot \vec v(T - t) \, \dd t \text.
\end{equation}
Under this product, the first-order time-derivative operator becomes symmetric:
\begin{equation}
    \inner{ \dot{\vec u}}{\vec v}_{\mathrm c} = \inner{\vec u}{\dot{\vec v}}_{\mathrm c} \text, 
\end{equation}
provided that initial conditions are homogeneous. This symmetry allows odd-order derivatives (e.g., damping terms) to appear naturally in the Euler-Lagrange equations. Despite this advantage, the approach remains limited to linear systems, and the convolution product is not positive-definite, meaning critical points of the action need not correspond to extrema. Moreover, as \citeauthor{Dargush2012a} and collaborators note \cite{Dargush2012, Dargush2012a}, homogeneous initial conditions remain restrictive for general IVPs. Instead, they propose a new action principle adopting both convolutions and fractional derivatives.

\subsubsection{Fractional derivatives}

Fractional derivatives, which generalize ordinary differentiation to non-integer orders, offer another pathway for formulating nonconservative action principles. By incorporating terms in the Lagrangian that depend on fractional time-derivatives of the dynamical variables \cite{Riewe1996, Riewe1997, Dargush2012, Dargush2012a}, the resulting Euler-Lagrange equations naturally produce terms with odd-order, which are characteristic of dissipative forces. This approach is motivated by the observation that if the Lagrangian depends on the $M$-th derivative, the corresponding Euler-Lagrange equations will involve derivatives up to order $2M$.  For example, to obtain a damping force proportional to $\dot{\vec q}$, one might include terms involving fractional derivatives of order $M=1/2$ in the Lagrangian.
Nevertheless, while the underlying insight is intuitive, fractional derivatives often introduce significant mathematical complexity since they are non-local operators. Furthermore, there is no universal definition of fractional derivative; several formulations exist---Riemann-Liouville, Caputo, Gr\"unwald-Letnikov, etc---each with distinct properties, advantages, and limitations. With no prevalent consensus, despite their potential, fractional derivative-based approaches remain an active area of research, and still suffer in terms of practical implementation.

\subsubsection{Contact mechanics}

From a Hamiltonian standpoint, ``contact'' geometry provides a higher-dimensional phase space (a contact manifold) in which dissipative processes behave as flows preserving a contact form \cite{Bravetti2017}. The contact manifold can be seen as an extension of the traditional symplectic phase space, where the extra dimension is not the time variable (as in the classical ``extended phase space approach''), but a non-trivial dynamical variable. This line of research generalises many concepts from symplectic mechanics, such as canonical transformations, generating functions, invariants of motion, and the contact Hamilton–Jacobi equations.

\subsubsection{Extra degrees of freedom}

Finally, one can circumvent Bauer’s corollary by expanding the system’s configuration space, and thus embedding it into a conservative framework of higher-dimension. A classical example is Bateman’s trick \cite{Bateman1931}: pair each dissipative degree of freedom with an auxiliary ``mirror'' variable that effectively absorbs the lost energy. Each mirror variable evolves according to an ``adjoint'' system; the combined setup conserves energy and fits within the traditional variational framework. In this sense, \citeauthor{Bateman1931} demonstrated that any set of linear ordinary differential equations can indeed be obtained from a single scalar action, provided additional degrees of freedom are introduced.
Eventually, this secondary adjoint system was shown to be the \emph{time-reversed} counterpart of the primary system \cite{Tonti1973, Dekker1981}, providing intuition into how the time-symmetry conditions of Noether's theorem can be bypassed. A very similar result can be achieved by extending the variables to the complex plane, where the primary and secondary variables are cast as complex conjugates of one another \cite[see][]{Kosyakov2007}.
Bateman's variational principle was eventually re-discovered in Hamiltonian form by \citeauthor{Morse1953} \cite{Morse1953}, and is often called the Bateman dual Hamiltonian \cite{Dekker1981, Um2002}.

Although Bateman’s approach was originally restricted to linear systems and continued to rely on boundary conditions (thus not resolving the mismatch between initial and boundary values), it proved influential in the development of more sophisticated strategies. Indeed, by the mid-twentieth century, a related framework had arisen in the context of quantum field theory. The \citeauthor{Schwinger1961}-\citeauthor{Keldysh1964}, ``closed-time-path'' or ``in-in'' formalism \cite{Schwinger1961, Keldysh1964} provides a path-integral description for studying out-of-equilibrium quantum systems. In the classical limit, the approach was applied to effective field-theories in GR \cite{Galley2009}.
Eventually, these insights led \citeauthor{Galley2013} and collaborators \cite{Galley2013, Galley2014} to successfully reframe the action principle within an IVP framework \cite{Galley2013, Galley2014}, thus enabling the description of nonconservative classical processes.
Similarly to the \citeauthor{Schwinger1961}-\citeauthor{Keldysh1964} formalism, Galley's framework introduced coupled variables that no longer represent backwards time evolutions as had been done in \citeauthor{Bateman1931}'s procedure. Instead, under application of what is termed the ``physical limit'', the coupled variables now evolve identically. Meanwhile, a single scalar action is constructed where the original variables are integrated forwards in time while the additional variables are integrated backwards; further coupling terms between the two introduce nonconservative effects. The advantage of these methods is that they require minimal changes to the traditional action principle. 

Despite its elegance and versatility, Galley’s original treatment leaves certain Lagrangian subtleties unaddressed and, more critically, does not fully develop a corresponding Hamiltonian framework. In this work, we adopt Galley’s variable‑doubling method, introduce key clarifications in the Lagrangian formulation, and extend the formalism into the Hamiltonian domain. 

In Sec.~\ref{sec:nonconservative-lagrangian-formulation}, we present the doubled-action formulation with new insights which include a detailed discussion of boundary conditions, the extraction of the physical limit, the partition of the action into conservative and nonconservative contributions, and a class of Lagrangian transformations which keep the motion invariant. 
In Sec.~\ref{sec:nonconservative-Hamiltonian-formulation}, we correct and complete the Legendre transform in Galley’s approach, constructing the nonconservative Hamiltonian structure. Finally, in Sec.~\ref{sec:inverse-variational-problem}, we tackle the inverse problem by deriving Lagrangians and Hamiltonians from arbitrary discrete initial-value equations, thereby demonstrating that any classical IVP admits a Hamiltonian representation. This perspective paves the way for the application of Hamiltonian perturbation methods to a variety of previously inaccessible nonconservative systems.

Specifically, in the context of post-Newtonian binary dynamics, it can enable a systematic treatment of the dissipative sector, seamlessly integrated into preexisting computations. In a companion paper, we demonstrate how the Lie perturbation method can be embedded within the present Hamiltonian framework and applied to explicitly solve the dynamics of nonconservative systems. Accordingly, we determine the full temporal evolution of $2.5$PN radiating binaries including both secular and oscillatory terms. This work can have important consequences in the computation of accurate gravitational wave templates in the context of future generation detectors.

\section*{Notation conventions}

We shall work primarily with geometric notation, whereupon bold italic symbols (e.g.\ $\vec u$) denote vectors, and bold straight symbols (e.g.\ $\mat M$) denote linear maps (matrices). When necessary, we shall adopt the subscript indices $i$, $j$ or $k$ $\in \{1, 2, \ldots, N \}$ for the components of $N$-vectors, sequences or matrices. 

For ease of notation, we shall work with these geometric objects in Euclidean space equipped with the canonical dot product, so that we do not need to differentiate between contravariant and covariant forms.
Hence, partial derivatives acting on scalars shall be interpreted as vectors, and partial derivatives acting on vectors shall be interpreted as matrices, according to the convention
\begin{equation}
    \mat M = \frac{\partial \vec u}{\partial \vec x} \implies M_{ij} = \frac{\partial u_j}{\partial x_i} \text,
\end{equation}
where derivatives are applied on the left.
The transpose operation shall be expressed via $\tr{(\argdot)}$.

We shall reserve the latin subscript indices running from $a, b, \ldots, h$ to indicate labelling of the doubled variables. These subscripts take values in $\{\labelA, \labelB\}$ or $\{+, -\}$ depending on the parametrization which is chosen.
A function simultaneously dependending on both variables shall be denoted with, for instance, $f(\pathAB{\vec q}) \equiv f(\pathA{\vec q}, \pathB{\vec q})$ or $f(\vec q_\pm) \equiv f(\vec q_+, \vec q_-)$. For equations where either $\{+,-\}$ or $\{\labelA, \labelB\}$ can be assumed, we shall write $f(\pathany{\vec q})$.

We shall adopt summation convention on repeated indices and path labels, unless explicitly stated.


\section{Nonconservative Lagrangian formulation} \label{sec:nonconservative-lagrangian-formulation}

\subsection{Initial value problem formulation via variable doubling}

Consider a Lagrangian system with $N$ degrees of freedom, described by a set of generalised coordinates $\vec q(t) \in \mathbb{R}^N$ and velocities $\dot{\vec q}(t) \in \mathbb{R}^N$ (i.e., $M=1$). Hamilton's principle states that the physical path $\vec q$ taken by the system over the time interval $[\ts, \te]$, between two specified states $\vec q(\ts) = \vec q_s$ and $\vec q(\te) = \vec q_e$, is the one for which the action functional $\act$ is stationary---namely $\delta \act [\vec q, \delta \vec q] = 0$ with
\begin{equation}
    \act[\vec q] = \int_{\ts}^{\te} \lagstd(\vec q, \dot{\vec q}, t) \, \dd t \text. \label{eq:action_conservative}
\end{equation}
As discussed, this standard boundary value formulation is insufficient for describing arbitrary nonconservative dynamics. Instead, the principle should be reformulated in a manner compatible with initial conditions.
In the Galley formalism \cite{Galley2013}, this is accomplished by doubling the degrees of freedom: one now has two variables, $\pathA{\vec q}$ and $\pathB{\vec q}$, each representing the same physical quantity as the original $\vec q$. While the two variables are allowed to differ in the action (``\textit{off-shell}''), their stationary (``\textit{on-shell}'') paths must evolve identically.

To show how this is achieved, let us begin by considering a conservative system evolving according to the principle \eqref{eq:action_conservative}. An equivalent double-variable action can be constructed from the integration of the conservative Lagrangian $\lagstd$ forward in time along $(\pathA{\vec q}, \pathA{\dot{\vec q}})$, and then backward in time along $(\pathB{\vec q}, \pathB{\dot{\vec q}})$:
\begin{align}
\act[\pathA{\vec q}, \pathB{\vec q}] &= \int_{\ts}^{\te} \lagstd(\pathA{\vec q}, \pathA{\dot{\vec q}}, t) \,\dd t + \int_{\te}^{\ts} \lagstd(\pathB{\vec q}, \pathB{\dot{\vec q}}, t) \,\dd t \nonumber \\
    &= \int_{\ts}^{\te} \!\left[ \lagstd(\pathA{\vec q}, \pathA{\dot{\vec q}}, t) - \lagstd(\pathB{\vec q}, \pathB{\dot{\vec q}}, t) \right] \dd t \text.
\end{align}
Variation of this action independently for $\pathA{\vec q}$ and $\pathB{\vec q}$ clearly leads to identical dynamics for the two variables, each evolving in accordance with the conservative principle \eqref{eq:action_conservative}. 

The interest now lies in further introducing into the integrand a term $\mathcal K$, which couples the two paths and can give rise to nonconservative dynamics. Namely, consider the more general action:
\begin{equation}
\act[\pathA{\vec q}, \pathB{\vec q}] = \int_{\ts}^{\te} \lagext(\pathA{\vec q}, \pathB{\vec q}, \pathA{\dot{\vec q}}, \pathB{\dot{\vec q}}, t) \, \dd t \text, 
\label{eq:act12}
\end{equation}
with extended Lagrangian:

\begin{align}
\lagext(\pathAB{\vec q}, \pathAB{\dot{\vec q}}, t) & = \lagstd(\pathA{\vec q}, \pathA{\dot{\vec q}}, t) - \lagstd(\pathB{\vec q}, \pathB{\dot{\vec q}}, t) \nonumber \\
& + \mathcal K(\pathAB{\vec q}, \pathAB{\dot{\vec q}}, t) \text.
\label{eq:lagext}
\end{align}
The additional term $\mathcal K$ can describe nonconservative processes: by coupling the forwards  and backwards paths, the time symmetry of Eq.\ \eqref{eq:action_conservative} is allowed to be broken dynamically, even when $\lagext$ is autonomous.
The manner of selecting an appropriate function $\mathcal K$ will depend on the physical problem: typical prescriptions might rely on symmetry arguments (as is often done in traditional Lagrangian mechanics); on reverse-engineering of the equations of motion (see Sec.~\ref{sec:inverse-variational-problem}); or on integrating-out or coarse-graining a subset of degrees-of-freedom from a larger closed system. A further discussion can be found in \cite[Section II.F]{Galley2014}.
Generally, one assumes that the dynamics generated by $\mathcal K$ cannot originate from two generalized potentials $U(\pathA{\vec q}, \pathA{\dot{\vec q}}, t) - U(\pathB{\vec q}, \pathB{\dot{\vec q}}, t)$; otherwise, these effects could simply be absorbed into the conservative Lagrangian by redefining $\lagstd \to \lagstd + U$. 
In Appendix~\ref{sec:Helmholtz}, we present a criterion---analogous to the Helmholtz condition in conservative action principles---that partitions $\lagext$ into conservative and nonconservative contributions. In particular, we show that the derivatives of $\mathcal K$ with respect to positions and velocities must ideally yield divergence-free vector fields when evaluated in the so-called physical limit condition $\pathA{\vec q} = \pathB{\vec q}$.

Since the two variables are constructed to represent the same physical path, a natural requirement on the action is to impose that it be invariant under their exchange, up to a multiplicative constant. This forces the action to be antisymmetric:
\begin{equation}
    \act[\pathA{\vec q}, \pathB{\vec q}] = - \act[\pathB{\vec q}, \pathA{\vec q}] \text,
\end{equation}
which implies that the coupling term must similarly obey
\begin{align}
     \mathcal K(\pathA{\vec q}, \pathB{\vec q}, \pathA{\dot{\vec q}}, \pathB{\dot{\vec q}}, t) = - \mathcal K(\pathB{\vec q}, \pathA{\vec q}, \pathB{\dot{\vec q}}, \pathA{\dot{\vec q}}, t) \text.
\end{align}
We call this the label-exchange (anti-)symmetry.

From the double Lagrangian $\lagext$, one can then define the nonconservative canonical momenta
\begin{subequations}\label{eq:nonconservative_momentum_AB}
    \begin{empheq}[left=\empheqlbrace]{align}
        \pathA{\vec \pi} & = \frac{\partial \lagext}{\partial \pathA{\dot{\vec q}}} = \frac{\partial \lagstd}{\partial \pathA{\dot{\vec q}}} + \frac{\partial \mathcal K}{\partial \pathA{\dot{\vec q}}} \, , \\
        \pathB{\vec \pi} & = - \frac{\partial \lagext}{\partial \pathB{\dot{\vec q}}} = \frac{\partial \lagstd}{\partial \pathB{\dot{\vec q}}} - \frac{\partial \mathcal K}{\partial \pathB{\dot{\vec q}}} \, .
    \end{empheq}
\end{subequations}
The first term on the right of each equation is a conservative contribution $\vec p_a = \partial \lagstd / \partial \dot{\vec q}_a$, for $a \in \{\labelA, \labelB\}$. Note that the sign reversal in the second equation is adopted for consistency with $\pathB{\vec \pi} = \pathB{\vec p}$ when $\mathcal K = 0$.


\begin{figure}
    \centering
    \includegraphics[scale=0.22]{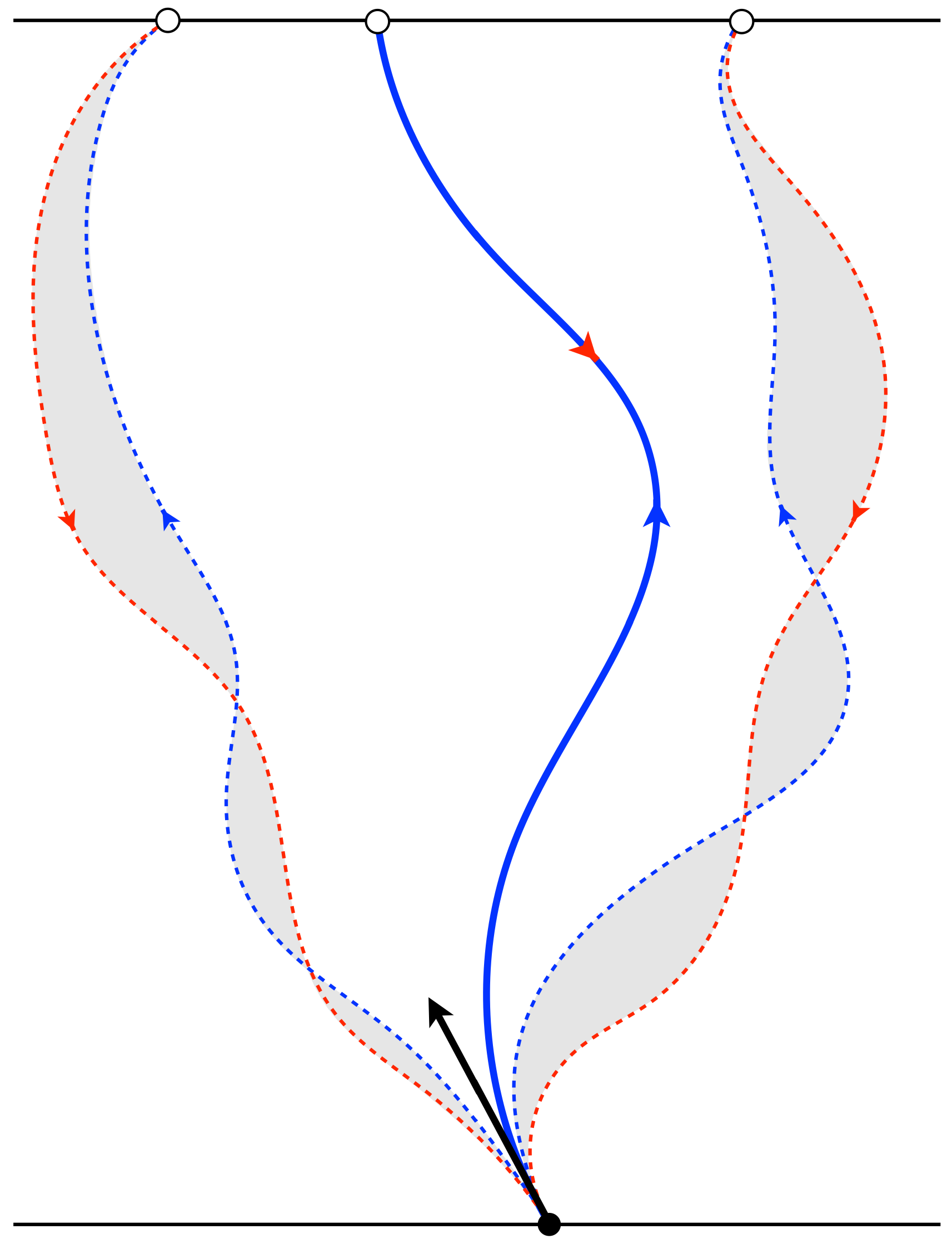}
    \setlength{\unitlength}{0.87cm}
    \begin{picture}(0,0)(-0.45,-0.0)
      \put(-7.95,0.4){\rotatebox{0}{$t=\ts$}}
      \put(-1.7,9.68){\rotatebox{0}{$t=\te$}}
      \put(-5.0,2.1){\rotatebox{0}{$\dot{\vec q}_s$}}
      \put(-3.7,0.4){\rotatebox{0}{$\vec q_s$}}
      \put(-4.5,6.1){\rotatebox{0}{$\pathA{\vec q}=\pathB{\vec q}$}}
    \end{picture}
    \caption{
    Illustration of the principle of stationary action formulated as an initial value problem via doubling of variables. The initial position and velocity at time $\ts$ are fixed (solid black circle and arrow), while the final state at time $\te$ remains free (white circles). The solid blue line denotes the physical trajectory that renders the action stationary, and the dashed lines represent the corresponding virtual displacements. The integration direction is indicated by colour: blue lines and arrows for forward integration and red lines and arrows for backward integration.
    }
    \label{fig:bvp_stationary_principle}
\end{figure}
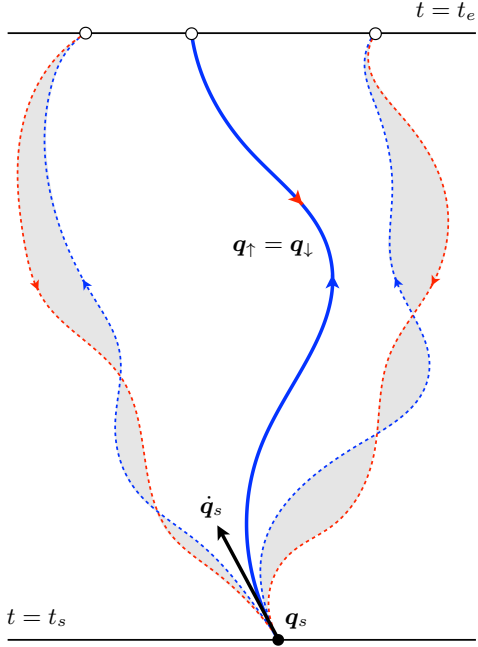

The action principle must also be accompanied by appropriate boundary conditions which lead to (i) the unique physical solution, and (ii) vanishing integration boundaries of the action variations. In order to establish a variational principle satisfying both points (i) and (ii), as many boundary conditions as required might be imposed on the variables $\pathA{\vec q}$ and $\pathB{\vec q}$, provided they are not in contradiction (which means that the variational problem is well-posed).

Consider a variation of the action \eqref{eq:act12} with respect to the two variables $\pathA{\vec q}$ and $\pathB{\vec q}$ independently:
\begin{align}
    \delta \act & = \left[ \delta_{ab} \, \delta \vec q_a \cdot \frac{\partial \lagext}{\partial \dot{\vec q}_b} \right]_{\ts}^{\te} \nonumber\\
    & + \int_{\ts}^{\te} \!\! \dd t \, \delta_{ab} \, \delta \vec q_a \cdot \left( \frac{\partial \lagext}{\partial \vec q_b} - \frac{\dd}{\dd t} \frac{\partial \lagext}{\partial \dot{\vec q}_b} \right) \, , \label{eq:nonconservative_action_variation}
\end{align}
where we introduce a Kronecker delta $\delta_{ab}$ summing along the path labels as stipulated in the notations section.
In order for the integration boundaries on the right-hand side (RHS) of Eq.\ \eqref{eq:nonconservative_action_variation} to vanish, a few choices are possible. At this step, we prefer the following conditions to the ones supplied by Galley in \cite{Galley2013}:\!\!
\begin{subequations}\label{eq:action_principle_initial_conditions}
\begin{empheq}[left=\empheqlbrace]{align}
  \pathA{\vec q}(\ts) & = \pathB{\vec q}(\ts) = \vec q_s \, , \\
  \pathA{\dot{\vec q}}(\ts) & = \pathB{\dot{\vec q}}(\ts) = \dot{\vec q}_s \, .
\end{empheq}
\end{subequations}
These equations enforce $4N$ conditions and impose that the path's variations vanish at the starting state---namely $\delta \vec q_a(\ts) = 0$ and $\delta \dot{\vec q}_a(\ts) = 0$, for $a \in \{\labelA, \labelB\}$. This choice is motivated by the intention of formulating a least-action principle as a genuine IVP. The boundary term in Eq.\ \eqref{eq:nonconservative_action_variation} thus reduces to
\begin{align}
\big[ \pathA{\vec \pi} \cdot \delta \pathA{\vec q} - \pathB{\vec \pi} \cdot \delta \pathB{\vec q} \big]_{\ts}^{\te} & = \pathA{\vec \pi} (\te) \cdot \delta \pathA{\vec q} (\te) \nonumber\\
 & - \pathB{\vec \pi} (\te) \cdot \delta \pathB{\vec q} (\te) \, .
\label{eq:boundaryremains}
\end{align}

As is required by item (ii), these boundary terms must vanish. A possibility for that is to impose the following $2N$ additional conditions:
\begin{subequations}\label{eq:action_principle_initial_conditions_te}
\begin{empheq}[left=\empheqlbrace]{align}
  \pathA{\vec q}(\te) & = \pathB{\vec q} (\te) \, , \label{eq:BCposte}\\
  \pathA{\dot{\vec q}}(\te) & = \pathB{\dot{\vec q}} (\te) \, . \label{eq:BCvitte}
\end{empheq}
\end{subequations}
Crucially, these boundary conditions do not fix the reference values of the final states, but merely require the two paths to coincide. This specification is entirely consistent with our lack of knowledge of the final states, beyond being a requirement for the variational formulation. 
Two immediate consequences arise. On the one hand, condition \eqref{eq:BCposte} leads to the equality between the two paths' variations at the end of integration, namely $\delta \pathA{\vec q} (\te) = \delta \pathB{\vec q} (\te)$. On the other hand, the anti-symmetry of the nonconservative Lagrangian under the $\pathinterchange$ label interchange, together with conditions \eqref{eq:action_principle_initial_conditions_te} combined, implies that $\pathA{\vec \pi} (\te) = \pathB{\vec \pi} (\te)$. Hence, the RHS in Eq.\ \eqref{eq:boundaryremains} vanishes, ensuring in this way the well-posedness of the variational principle as established below.

After setting $\delta \act = 0$, one derives the following pair of Euler-Lagrange equations:
\begin{equation}
\frac{\partial \lagext}{\partial \vec q_a} - \frac{\dd}{\dd t} \left( \frac{\partial \lagext}{\partial \dot{\vec q}_a} \right) = 0 \, , \quad \text{for } a \in \{\labelA, \labelB\}.
\label{eq:EL_nonconservative}
\end{equation}
For each label $a$, the corresponding equation involves both paths $\pathA{\vec q}$ and $\pathB{\vec q}$, as well as their time derivatives. Because $\lagext$ is antisymmetric under the label exchange, the two equations are similarly related by the simultaneous swap
\begin{equation}
    (\pathA{\vec q}, \pathA{\dot{\vec q}}, \pathA{\ddot{ \vec q}}) \longleftrightarrow (\pathB{\vec q}, \pathB{\dot{\vec q}}, \pathB{\ddot{ \vec q}})
\end{equation}
up to an overall (irrelevant) minus sign.

To extract physically meaningful dynamics from this doubled system, it is practical to reduce the pair to a single equation. Suppose the pair of equations meets the usual conditions of, for instance, the Picard–Lindel\"of theorem---namely, they can be cast into a system of first-order ordinary differential equations (ODEs) in standard form, with Lipschitz-continuous right-hand sides. In that case, the IVP admits a unique solution.
Given that we supply identical initial conditions for both paths $a = \labelA$ and $a = \labelB$ [cf.\ Eqs.\  \eqref{eq:action_principle_initial_conditions}], it follows from the symmetries that $\pathA{\vec q} = \pathB{\vec q}$ produces a \emph{valid} solution for the system.
Uniqueness then \emph{guarantees} that this property is satisfied. To see this, suppose a solution existed where $\pathA{\vec q}(t) \neq \pathB{\vec q}(t)$ for some time $t$. Interchanging the $\pathinterchange$ labels in this solution would generate a distinct solution pair satisfying both \eqref{eq:EL_nonconservative} and the original initial conditions---a contradiction. Consistency with the uniqueness property thus necessitates the functional equality $\pathA{\vec q} = \pathB{\vec q}$. The trajectories are consequently confined to the sub-manifold defined by this property, whereupon the double-path system collapses into a single physical path.

Hereafter, we shall call the identification:
\begin{equation}\label{eq:PL}
\begin{aligned}
\pathA{\vec q}(t) &= \pathB{\vec q} (t) \equiv \vec q (t) \, , \\
\pathA{\dot{\vec q}}(t) &= \pathB{\dot{\vec q}} (t) \equiv \dot{\vec q} (t) \, , \\
\pathA{\ddot{ \vec q}}(t) &= \pathB{\ddot{ \vec q}} (t) \equiv \ddot{\vec q} (t) \, 
\end{aligned}
\end{equation}
the \emph{physical limit}, valid on-shell for any time $t$. 
More precisely, we define a sub-manifold $\Sigma$ as the constrained space of smooth class $\mathcal C_2$ solutions
\begin{equation}
\Sigma=\big\{ (\pathA{\vec q}, \pathB{\vec q}) \in \mathcal C_2([\ts, \te], \mathbb{R}^{2N}) ; \pathA{\vec q} = \pathB{\vec q} \equiv \vec q \big\} \, .
\end{equation}
In short, we see that the physical limit emerges naturally as a consequence of (i) the anti-symmetry of $\lagext$ and (ii) the re-formulation of the least-action principle as an IVP; it is not an artificially imposed constraint, contrary to what one might infer from \cite{Galley2013}. Accordingly, most properties of standard action principles carry on to the nonconservative principle; the physical limit is merely a simplification device deduced from the resulting Euler-Lagrange equations.
We shall subsequently adopt the notation $(\argdot )_{\PL\!}$
to denote some expression evaluated in the physical limit.
By virtue of the identification \eqref{eq:PL}, the conservative nature of Lagrangian systems is broken.
Figure~\ref{fig:bvp_stationary_principle} summarizes the action principle integration paths and boundary conditions.

A direct consequence of the physical limit solution is that one of the Eqs.\ \eqref{eq:EL_nonconservative} becomes redundant and, after substituting for $\lagext$ [Eq.\ \eqref{eq:lagext}] into Eq.\ \eqref{eq:EL_nonconservative}, the dynamics simplifies to the single physical Euler-Lagrange equation
\begin{align}
\frac{\partial \lagstd}{\partial \vec q} - \frac{\dd}{\dd t} \frac{\partial \lagstd}{\partial \dot{\vec q}} = \frac{1}{2}\left[ \frac{\partial \mathcal K}{\partial \pathB{\vec q}} - \frac{\partial \mathcal K}{\partial \pathA{\vec q}} - \frac{\dd}{\dd t} \left(  \frac{\partial \mathcal K}{\partial \pathB{\dot{\vec q}}} 
- \frac{\partial \mathcal K}{\partial \pathA{\dot{\vec q}}}
\right) \right]_{\PL} \!\text. 
\end{align}
For convenience, we have anti-symmetrized the derivatives on the RHS with respect to $\pathinterchange$ label interchange, altough this is not required. The terms inside the square brackets contribute as nonconservative generalized forces. 

\subsection{Plus-minus variables}

It is often convenient to recast the framework into the $(+,-)$ or Keldysh variables \cite{Keldysh1964}, which describe the average path and their relative deviation:
\begin{align}
    \vec q_+ = \frac{\pathA{\vec q} + \pathB{\vec q}}{2} \text, && \vec q_- = \pathA{\vec q} - \pathB{\vec q} \text. \label{eq:pm_variables_q}
\end{align}
The physical limit becomes $\vec q_+ \to \vec q$ and $\vec q_- \to 0$. Accordingly, we shall often refer to $\vec q_+$ as the physical variable and to $\vec q_-$ as the virtual variable---since, in the action, $\vec q_-$ effectively behaves as a virtual displacement. 
These definitions are understood as functional transformations on the entire time paths, so that analogous relations apply to all higher-order time derivatives. In a similar manner, the momenta will be transformed as:
\begin{align}
    \vec \pi_+ &= \frac{\pathA{\vec \pi} + \pathB{\vec \pi}}{2} = \dext{\lagext}{\dot{\vec q}_-} \text, & \vec \pi_- &= \pathA{\vec \pi} - \pathB{\vec \pi} = \dext{\lagext}{\dot{\vec q}_+} \text, \label{eq:pm_variables_p}
\end{align}
where the Lagrangian must now be seen as a function of $(\vec q_\pm, \dot{\vec q}_\pm)$ by inversion of the relations \eqref{eq:pm_variables_q}.
More compactly, the momenta can be described via the labels introduced in the notation section:
\begin{align}
    \vec \pi_a = \pathmetric \dext{\lagext}{\dot{\vec q}_b} 
    \text.  \label{eq:nonconservative_momentum}
\end{align}
This label notation allows a unified representation of equations through either variable set $(\labelA, \labelB)$ or $(+, -)$, whereupon the Einstein summation convention is adopted. The ``path label metric'' $\pathmetric$, expressed in the $(+, -)$ variable set, has matrix representation $[\pathmetric] = \antidiag(1, 1)$, which when replaced in Eq.~\eqref{eq:nonconservative_momentum} yields Eqs.~\eqref{eq:pm_variables_p}. Meanwhile, when expressed in the $(\labelA, \labelB)$ parametrization, the metric coefficients are $[\pathmetric] = \diag(1, -1)$, and we recover Eqs.~\eqref{eq:nonconservative_momentum_AB}.

We now return to the boundary conditions introduced in the previous section to elucidate a few additional aspects. In the $(+, -)$ parametrization, the $6N$ variational conditions can be more cleanly decoupled; they are expressed as:
\begin{subequations}\label{eq:boundary_conditions_pmv}
\begin{align}
    &\vec q_+(\ts) = \vec q_s \text, && \dot{\vec q}_+(\ts) = \dot{\vec q}_s \text, \label{eq:boundary_conditions_pmv_+} \\
    &\vec q_-(\ts) = 0 \text, && \dot{\vec q}_-(\ts) = 0 \text, \label{eq:boundary_conditions_pmv_-_ti} \\
    &\vec q_-(\te) = 0 \text, && \dot{\vec q}_-(\te) = 0 \text. \label{eq:boundary_conditions_pmv_-_tf} 
\end{align}    
\end{subequations}
Firstly, the equality conditions at initial time [Eq.~\eqref{eq:boundary_conditions_pmv_-_ti}] and final time [Eq.~\eqref{eq:boundary_conditions_pmv_-_tf}] are clearly redundant, describing Cauchy data for the same degree of freedom $\vec q_-$, at different times $\ts$ and $\te$. From this lens, these two sets of conditions can independently lead to the physical limit constraint $\vec q_-(t) = 0$, via either forward or backward integration of the Euler-Lagrange equations [Eqs.\ \eqref{eq:EL_nonconservative}]. However, the presence of redundant conditions is not an issue because they are mutually compatible and do not over-constrain the solution---that is, all four relations in [Eqs.~(\ref{eq:boundary_conditions_pmv_-_ti}, \ref{eq:boundary_conditions_pmv_-_tf})] are satisfied in the physical limit, and no contradictions arise. 
Secondly, appropriate initial conditions for the physical path $\vec q_+$ are vital for the IVP to be completely and uniquely formulated. We call any such conditions the \emph{physical conditions}. Because our main goal is to ensure a genuine, well-posed \emph{initial value} action principle, the physical conditions will in general be supplied via initial data, as indeed is fulfilled through Eq.~\eqref{eq:boundary_conditions_pmv_+}. However, we remark that the double-variable formulation does also allow other forms of physical conditions (Dirichlet, Neumann, mixed, etc) which might be more appropriate for certain physical scenarios; in such cases, Eq.~\eqref{eq:boundary_conditions_pmv_+} can be freely adapted.
Lastly, different combinations of conditions can achieve the vanishing of the integration boundaries of the action variations. For example, this is achieved by the two pairs of equality conditions alone [Eqs.~(\ref{eq:boundary_conditions_pmv_-_ti}, \ref{eq:boundary_conditions_pmv_-_tf})], which justifies our choice to adopt them both, despite some redundancy. Galley’s choice differs from ours, and does not decouple so cleanly---it is discussed below. 
In general, a variant of the final-time equality condition [Eq.~\eqref{eq:boundary_conditions_pmv_-_tf}] is crucial for the vanishing of the boundaries in an IVP, since the undesirable alternative would be to fix final-time data for the physical variables $\vec q_+$.

Another feasible variation substitutes the conditions on velocities for conditions on momenta. When the Lagrangian is sufficiently regular, the momenta can be inverted and the two are equivalent.
For instance, on one hand, in \cite{Galley2013}, Galley appoints the variational conditions
\begin{subequations}
\begin{align}
    &\pathA{\vec q}(\ts) = \pathA[,s]{\vec q} \text, &
    &\pathB{\vec q}(\ts) = \pathB[,s]{\vec q} \text, \\
    &\vec q_-(\te) = 0 \text, &
    &\dot{\vec q}_-(\te) = 0 \text.
\end{align}
\end{subequations}
On the other hand, in \cite{Galley2014}, $\dot{\vec q}_-(\te) = 0$ is replaced by $\vec \pi_-(\te) = 0$. For regular Lagrangians, this choice should make no difference.
Notice that the physical initial conditions appointed by Galley are only for the position (i.e., $\vec q_+(\ts) = (\pathA[,s]{\vec q} + \pathB[,s]{\vec q})/2$), while ensuring a unique Euler-Lagrange solution requires additional initial data for the velocity or momentum (fixing of $\dot{\vec q}_+(\ts)$ or $\vec \pi_+(\ts)$). 

We would also like to note that in Galley's formulation, strictly speaking, the non-physical condition $\vec q_-(\ts) = \pathA[,s]{\vec q} - \pathB[,s]{\vec q} \neq 0$ was not forbidden a priori. However, if such inequality were to take place, a contradiction arises and the problem becomes ill-posed, since the final-time conditions provide Cauchy data at $\te$ which fix the trajectory as the constant $\vec q_-(t) = 0$ for all $t$. Of course, the condition $\vec q_-(\ts) \neq 0$ is moreover non-physical and will never occur in practice. We consequently believe that a more appropriate implementation would directly restrict $\vec q_-(\ts) = 0$ when specifying the variational principle.

In the $(+, -)$ variable set, the variations of the action can be directly performed with respect to $(\vec q_+, \vec q_-)$, leading to Euler-Lagrange equations which keep the same general form [Eq.~\eqref{eq:EL_nonconservative}]; however, the label $a$ now takes values in $\{ +, -\}$. In the physical limit, the condition $a = +$ reduces to a trivial identity ($0 = 0$), while $a = -$ provides the physically meaningful equation:
\begin{align}
\frac{\partial \lagstd}{\partial \vec q} - \frac{\dd}{\dd t} \frac{\partial \lagstd}{\partial \dot{\vec q}} = - \left( \dext{\mathcal K}{\vec q_-} - \frac{\dd}{\dd t} \dext{\mathcal K}{\dot{\vec q}_-} \right)_{\!\PL} . \label{eq:EL_nonconservative_physical}
\end{align}

As Galley has emphasized, it is often advantageous (but not necessary) to expand the Lagrangian $\lagext$ perturbatively in powers of $\vec q_- = \pathA{\vec q} - \pathB{\vec q}$ and $\dot{\vec q}_- = \pathA{\dot{\vec q}} - \pathB{\dot{\vec q}}$ around the origin. For convenience, we introduce a shorthand Landau notation:
\begin{equation}
    \vorder{n} = \mathcal O(\lVert (\vec q_-, \dot{\vec q}_- )\rVert ^n) \text.
\end{equation}
Because the Lagrangian is anti-symmetric under $\pathinterchange$ label exchange, only odd perturbative powers appear in this expansion. Furthermore, any terms which are of asymptotic order $\vorder{2}$ in the Lagrangian vanish in the physical limit Euler-Lagrange equation [Eq.\ \eqref{eq:EL_nonconservative_physical}], rendering them irrelevant to the physical dynamics. 
To illustrate, consider a nonconservative component prescribed in the form
\begin{align}
    \mathcal K(\vec q_\pm, \dot{\vec q}_\pm, t) &= \vec q_- \cdot \mathcal K_{\vec q}(\vec q_+, \dot{\vec q}_+, t) \nonumber\\ 
    &\, + \dot{\vec q}_- \cdot \mathcal K_{\dot{\vec q}}(\vec q_+, \dot{\vec q}_+, t) + \vorder{3} \text,
\end{align}
where the functions $\mathcal K_{\vec q}$ and $\mathcal K_{\dot{\vec q}}$ may be chosen arbitrarily. Substituting into Eq.~\eqref{eq:EL_nonconservative_physical} then yields
\begin{equation}
    \frac{\partial \lagstd}{\partial \vec q} - \frac{\dd}{\dd t} \frac{\partial \lagstd}{\partial \dot{\vec q}} = - \left( \mathcal K_{\vec q} + \frac{\dd}{\dd t} \mathcal K_{\dot{\vec q}} \right) \, \text.
\end{equation}
The terms of order $\vorder{3}$ in $\mathcal K$ do not contribute since their derivatives are of order $\vorder{2}$ and thus vanish in the physical limit. Because the functions $\mathcal K_{\vec q}$ and $\mathcal K_{\dot{\vec q}}$ can be chosen freely, this framework is capable of describing virtually any nonconservative force on the RHS.

Consequently, it is often convenient to linearize the action with respect to $(-)$ terms. The process does not incur any approximation but simply benefits from the redundant structure of the doubled variables to isolate the physically relevant terms. In the following section, we derive a new perspective from which the linearized action can be analyzed.

\subsection{Linearized action}\label{sec:linear_act}

In the discussion that follows, we shall be systematically discarding terms of order $\vorder{2}$. We begin by considering the double variable action, linearized around the physical limit:
\begin{equation}
\act[\vec q_+, \vec q_-] = 
\int_{\ts}^{\te} \!\! \dd t \, \bigg\{ \dot{\vec q}_- \cdot \left( \,\dext{\lagext}{\dot{\vec q}_-} \right)_{\!\PL\!}\! + \vec q_- \cdot \left( \,\dext{\lagext}{\vec q_-} \right)_{\!\PL\!}  \bigg\}  \text, \label{eq:linearized_action}
\end{equation}
with expressions enclosed in $(\argdot)_{\PL\!}$ evaluated on the plus variables. 
From this standpoint, when deriving the physical Euler-Lagrange equations [Eq.\ \eqref{eq:EL_nonconservative_physical}], $\vec q_-$ effectively acts as a virtual displacement, since integration by parts directly yields
\begin{equation}
    \delta \act = \int_{\ts}^{\te} \! \delta\vec q_- \cdot \left( \,\dext{\lagext}{\vec q_-} - \frac{\dd}{\dd t} \dext{\lagext}{\dot{\vec q}_-} \right)_{\!\PL\!} \,\dd t \text,
\end{equation}
and the piece multiplying $\delta\vec q_-$ is indeed the equation of motion.

We find it instructive to introduce a differential operator, representing a Lie derivative along the flow of the dynamical trajectories and time. We define the ``on-shell time derivative'' by
\begin{equation}
    \Dt = \frac{\partial}{\partial t} + \delta_{ab} \left( \dot{\vec q}_a \cdot \dext{}{\vec q_b} + \vec U_a \cdot \dext{}{\dot{\vec q}_b} \right) \text. \label{eq:onshell_dt}
\end{equation}
The vector fields $\vec U_a = \vec U_a\big(\pathany{\vec q}, \pathany{\dot{\vec q}}, t\big)$ play the role of on-shell accelerations, so that $\ddot{\vec q}_a = \vec U_a$ when the Euler-Lagrange equations hold. In other words, $\Dt$ corresponds to a total time derivative where accelerations are replaced by corresponding vector fields generated by the dynamics---depending only on positions, velocities and time. By construction, the operator $\Dt$ satisfies the off-shell version of the Euler-Lagrange equations [cf.\ Eq.\ \eqref{eq:EL_nonconservative}]:
\begin{equation}
    \Dt[\vec \pi_a] = \pathmetric[ab] \dext{\lagext}{\vec q_b}. \label{eq:onshell_dp_dt}
\end{equation}

The explicit expressions for $\vec U_a$ can be obtained from Eq.~\eqref{eq:onshell_dp_dt} assuming a sufficiently regular Lagrangian.
Expanding the LHS of Eq.~\eqref{eq:onshell_dp_dt} for $a \in \{+,-\}$ allows one to recover a block-matrix linear equation for $\vec U_\pm$, namely
\begin{equation}
\begin{pmatrix}
    \mat H_{-+} && \mat H_{--} \\
    \mat H_{++} && \mat H_{+-}
\end{pmatrix}
\begin{pmatrix}
    \vec U_+ \\
    \vec U_-
\end{pmatrix}
=
\begin{pmatrix}
    \vec \chi_+ \\
    \vec \chi_-
\end{pmatrix},\label{eq:G_mat_solve}
\end{equation}
where $\mat H_{ab}$ are Hessians of $\lagext$ with respect to velocities
\begin{equation}
  \mat H_{ab} = \frac{\partial^2 \lagext}{\partial\dot{\vec q}_a \partial\dot{\vec q}_b\!}  = \pathmetric[bc] \dext{\vec \pi_c}{\dot{\vec q}_a\,\,} \text,
\end{equation}
and $\vec \chi_a$ are given by the difference
\begin{equation}
    \vec \chi_a = \pathmetric \dext{\lagext}{\vec q_b} - \delta_{bc} \dot{\vec q}_b \cdot \frac{\partial \vec \pi_a}{\partial \vec q_c\!} - \frac{\partial \vec\pi_a\!}{\partial t} \,\text.
\end{equation}
In the linear action formulation, $\mat H_{--}$ vanishes, as can be deduced from the expression of $\Lambda$ in Eq.~\eqref{eq:linearized_action}. The physically relevant Hessian is $\mat H_{-+} = \tr{\mat H_{+-}}$, which relates to the dynamics of $\vec q_+$ [we shall see in Eq.\ \eqref{eq:ode_2nd_order_hessian}].  
Similarly to standard classical mechanics, the Lagrangian $\lagext$ is assumed to be sufficiently regular, so that in the physical limit $\mat H_{-+}$ is non-singular and its inverse is locally well-defined.
Even in the non-linear formulation, Eq.~\eqref{eq:G_mat_solve} can be typically solved by means of block-matrix inversion with respect to the $\mat H_{-+}$ Schur complement. 
Solving explicitly for the linear case, we obtain
\begin{subequations}\label{eq:on_shell_dt_G}
\begin{empheq}[left=\empheqlbrace]{align}
    \vec U_+ &= \mat H_{-+}^{-1} \, \vec \chi_+, \\
    \vec U_- &= \mat H_{+-}^{-1} \, \vec \chi_- - \mat H_{+-}^{-1} \mat H_{++} \mat H_{-+}^{-1} \, \vec \chi_+.
\end{empheq}
\end{subequations}

The next step is to reframe the linearized action [Eq.~\eqref{eq:linearized_action}] in terms of the momentum function and the $\Dt$ operator, which yields the convenient form
\begin{align}
\act[\vec q_+, \vec q_-] 
&= \int_{\ts}^{\te} \!\Big( \dot{\vec q}_- \cdot \vec \pi_+ + \vec q_- \cdot \Dt[\vec \pi_+] \Big) \,\dd t \nonumber\\
&= \int_{\ts}^{\te} \Dt[\vec q_- \cdot \vec \pi_+] \,\dd t \text,\label{eq:linearized_action_Lie_form}
\end{align}
which, remarkably, represents a time integral of the Lie derivative along the dynamics. We emphasize that this integral does not evaluate to a boundary term, since the Lie operator $\Dt$ differs from a traditional total time derivative $\dd/\dd t$.
In this linear action, the $(+)$ momentum is exactly the physical momentum $\vec\pi_+ = \vec\pi(\vec q_+, \dot{\vec q}_+, t)$, so that $\vec U_-$, which is linked to the evolution of $\vec q_-$, does not contribute in Eq.~\eqref{eq:linearized_action_Lie_form}, as can be verified by re-expanding the operator $\Dt$.

Upon varing $\act$ with respect to $\vec q_-$ one now recovers
\begin{equation}
    \frac{\dd \vec \pi_+\!}{\dd t} = \Dt[\vec \pi_+] \text,
\end{equation}
which is precisely the equation of motion. Namely, the operators $\dd/\dd t$ and $\Dt$ can both be expanded to yield
\begin{equation}
    \mat{H}_{-+} \left( \ddot{\vec q}_+ - \vec U_+ \right) = 0 \text. \label{eq:ode_2nd_order_hessian}
\end{equation}
Applying the physical limit, since $\mat{H} = \left( \mat{H}_{-+} \right)_{\PL\!}$ is (locally) invertible, the physical equation is recovered in standard form
\begin{equation}
    \ddot{\vec q} = \vec U(\vec q, \dot{\vec q}, t) \text, \label{eq:ode_2nd_order}
\end{equation}
with $\vec U(\vec q, \dot{\vec q}, t) = \left( \vec U_+ \right)_{\PL}$.

The entire dynamical content of the linearized action is thus encoded in the Lie‐derivative operator $\Dt$, and parametrized solely by the vector field $\vec U_+$. Importantly, the remaining vector field $\vec U_-$, associated with the evolution of the virtual variable $\vec q_-$, does not contribute to the physics.  The operator $\Dt$ then acts on the product $\vec q_- \cdot \vec \pi_+$ [c.f.\ Eq.~\eqref{eq:linearized_action_Lie_form}], which contains only information about the definition of the momentum. 
In other words, the momentum map $(\vec q_+, \dot{\vec q}_+) \mapsto \vec \pi_+$ can be prescribed independently to its dynamics, which remains entirely contained in $\Dt$.

In this way, the physical content of the theory remains invariant under arbitrary shifts of $\vec \pi_+$. This amounts to gauge-like transformations in the momentum space. As long as the resulting $\vec \pi_+$ is a local diffeomorphism (so that $\mat H_{-+}$ stays invertible), the second-order equation [Eq.~\eqref{eq:ode_2nd_order_hessian}] is left unchanged. For instance, the straightforward choice $\vec \pi_+ = \dot{\vec q}_+$ with Lagrangian $\lagext = \Dt[\vec q_- \cdot \dot{\vec q}_+]$
can in general deliver exactly the same equations of motion as any other well‐posed momentum prescription.

\subsection{Canonical gauge shift}\label{sec:linear:gauge_shift}

We now consider explicitly the effect of such a momentum shift in the linearized Lagrangian, which we shall henceforth refer to as a ``canonical gauge shift''. A full demonstration in the non-linearized formulation is also provided in Sec.~\ref{sec:canonical_gauge_shift}, where we see that indeed this is a feature of the general nonconservative framework. We begin by considering a Lagrangian transformation of the form 
\begin{equation}
    \lagext \to \lagext' = \lagext + \Dt \big[\vec q_- \cdot \vec \phi \big] = \Dt' \big[\vec q_- \cdot (\vec \pi_+ + \vec \phi)\big] \text,
\end{equation}
for an arbitrary vector field $\vec \phi$ which is function of the $(+)$ variables and time, under the constraint that $\lagext'$ be regular. Importantly, this transformation preserves the physical equations of motion, leaving the physical acceleration $\vec U_+$ unchanged. To see this, consider the on-shell time derivative of the transformed momentum, which satisfies by construction
\begin{equation}
    \Dt'[\vec \pi_a'] = \pathmetric[ab] \dext{\lagext'\!}{\vec q_b\,} \,. \label{eq:onshell_dp_dt_gauge}
\end{equation}
Expanding for $a = +$, one can see that all corrections cancel at $\vorder{2}$ and hence $\vec U_+$ is not affected by the momentum shift. Only $\vec U_-$ is modified, which is indeed a requirement so that the Lagrangian structure is retained.

The transformed momenta are likewise derived from $\lagext'$; in the $(+,-)$ parametrization they equal
\begin{subequations}\label{eq:pi_gauge}
\begin{empheq}[left=\empheqlbrace]{align}
    \vec \pi_+' &= \dext{\lagext'\!}{\dot{\vec q}_-} = \vec \pi_+ + \vec \phi , \\
    \vec \pi_-' &= \dext{\lagext'}{\dot{\vec q}_+} = \vec \pi_- + \vec q_- \cdot \dext{\Dt[\vec\phi]}{\dot{\vec q}_+} + \dot{\vec q}_- \cdot \dext{\vec \phi}{\dot{\vec q}_+}.
\end{empheq}
\end{subequations}

\section{Nonconservative Hamiltonian formulation} \label{sec:nonconservative-Hamiltonian-formulation}

We are now ready to determine the general nonconservative Hamiltonian formulation via Legendre transformation of $\Lambda$.
As in conservative mechanics, a necessary and sufficient condition for the conversion is that the Legendre transform be a local diffeomorphism
. Specifically, the physical momentum $\vec \pi_+ = \partial \lagext/\partial \dot{\vec q}_-$ must have a local inverse with respect to the velocity $\dot{\vec q}_+$, which is required for computing the Hamiltonian function. From the inverse function theorem, this equates to the Lagrangian $\lagext$ being regular---that is, the Hessian of $\lagext$ with respect to $(+,-)$ velocities being non-degenerate in the physical limit:
\begin{equation}
    \det \mat H = \det \left( \frac{\partial^2 \lagext}{\partial \dot{\vec q}_+ \partial\dot{\vec q}_-} \right)_{\!\PL} \neq 0 \text.
\end{equation}

Whenever the inversion of $\vec \pi_+$ is instead a \emph{global} property, this allows the Legendre transform to be well-posed throughout the whole domain, and the equations of motion will be globally expressible in standard form $\ddot{\vec q} = \vec U(\vec q, \dot{\vec q}, t)$. This stronger property shall be assumed in Sec.~\ref{sec:inverse-variational-problem}, where we reconstruct a Lagrangian and Hamiltonian from a given ODE. In the absence of this property, one must be content with a more local description with coordinate charts.

The Hamiltonian is defined as a standard Legendre transform of $\lagext$ with momenta $\vec \pi_a$ prescribed by Eq.~\eqref{eq:nonconservative_momentum}:
\begin{equation}
    \hamext(\pathany{\vec q}, \pathany{\vec \pi}, t) 
    = \pathmetric \, \vec \pi_a \cdot \dot{\vec q}_b - \lagext(\pathany{\vec q}, \pathany{\dot{\vec q}}, t) \, ,
\label{eq:nonconservative_Legendre_transform}
\end{equation}
where the $\pathmetric$ factor is picked up from the momenta definitions.
As usual, the velocities $\dot{\vec q}_a$ should be understood as functions of the phase space variables $(\vec q_d, \vec \pi_d)$ obtained by inversion of the generalized momentum equations [Eqs.~\eqref{eq:nonconservative_momentum}].
To derive the Hamiltonian dynamics, we follow the usual canonical procedure, starting from the total differential of the Lagrangian and Hamiltonian functions:
\begin{subequations}
\begin{align}
    \dd \lagext &= \delta_{ab}\, \dext{\lagext}{\vec q_a} \cdot \dd \vec q_b + \delta_{ab}\, \dext{\lagext}{\dot{\vec q}_a} \cdot  \dd \dot{\vec q}_b + \frac{\partial \lagext}{\partial t} \dd t \text, \\
    \dd \hamext &= \delta_{ab}\, \dext{\hamext}{\vec q_a} \cdot \dd \vec q_b + \delta_{ab}\, \dext{\hamext}{\vec \pi_a} \cdot \dd \vec \pi_b + \frac{\partial \hamext}{\partial t} \dd t  \text. 
\end{align}    
\end{subequations}
The expression of the Legendre transform [Eq.\ (\ref{eq:nonconservative_Legendre_transform})] relates the two differentials via
\begin{equation}
    \dd \hamext = \pathmetric \, \dot{\vec q}_a \cdot \dd \vec \pi_b + \pathmetric \, \vec \pi_a \cdot \dd \dot{\vec q}_b - \dd \lagext \text.
\end{equation}
Replacing $\dd \lagext$ and $\dd \hamext$, cancelling out momentum terms via Eq.~\eqref{eq:nonconservative_momentum}, and identifying the coefficients yields Hamilton's equations:
\begin{subequations}\label{eq:ham_eqs}
\begin{empheq}[left=\empheqlbrace]{align}
  \dot{\vec \pi}_a & = \pathmetric  \dext{\lagext}{\vec q_b} = - \pathmetric \dext{\hamext}{\vec q_b} \text, \label{eq:ham_eqs_leg_tr_pi}\\
  \dot{\vec q}_a & = \pathmetric \dext{\hamext}{\vec \pi_b} \text,
\end{empheq}
\end{subequations}
and
\begin{equation}
    \frac{\partial \hamext}{\partial t} = - \frac{\partial \lagext}{\partial t} \text,
\end{equation}
where in Eq.\ \eqref{eq:ham_eqs_leg_tr_pi} we have substituted in the Euler-Lagrange equations [Eq.~\eqref{eq:EL_nonconservative}]. 

An alternative form of the nonconservative Hamiltonian can allow a more explicit identification of the conservative and nonconservative contributions. Consider the conservative Hamiltonian $\hamstd$, obtained from the usual Legendre transform of $\lagstd$:
\begin{equation}
    \lagstd(\vec q_a, \dot{\vec q}_a, t) = \vec p_a \cdot \dot{\vec q}_a - \hamstd(\vec q_a, \vec p_a, t) \text,
\end{equation}
for $a \in \{\labelA, \labelB\}$. We precise that the above equation is only valid in the $(\labelA, \labelB)$ variable set. We also emphasize the presence of the conservative momentum $\vec p_a$ appearing on the RHS. Substituting for $\lagstd$ from the previous equation into Eq.\ \eqref{eq:nonconservative_Legendre_transform} evaluated in the $(\labelA, \labelB)$ variables leads to:
\begin{align}
    \hamext(\pathany{\vec q}, \pathany{\vec \pi}, t) & = 
    \hamstd\Big(\pathA{\vec q}, \pathA{\vec \pi} - \frac{\partial \mathcal K}{\partial \pathA{\dot{\vec q}}} , t\Big) \nonumber \\
    & - \hamstd\Big(\pathB{\vec q}, \pathB{\vec \pi} + \frac{\partial \mathcal K}{\partial \pathB{\dot{\vec q}}}, t \Big) + \mathcal R(\pathany{\vec q}, \pathany{\vec \pi}, t) \text,\label{eq:hamext}
\end{align}
where the conservative momenta $\vec p_a$ have been replaced by their expressions in terms of nonconservative variables.
Meanwhile, $\mathcal R$ holds nonconservative contributions:
\begin{align}
    \mathcal R(\pathany{\vec q}, \pathany{\vec \pi}, t) & = \delta_{ab} \, \dot{\vec q}_a \cdot \dext{\mathcal K}{\dot{\vec q}_b} (\pathany{\vec q}, \pathany{\vec \pi}, t) \nonumber\\ 
    & - \mathcal K(\pathany{\vec q}, \pathany{\vec \pi}, t) \text.
\end{align}
We highlight that $\mathcal R$ is not the Legendre transformation of $\mathcal K$ since $\dot {\vec q}_a = \dot {\vec q}_a(\pathany{\vec q}, \pathany{\vec \pi}, t)$ are obtained via inversion of the momenta of $\lagext$ [Eq.\ \eqref{eq:nonconservative_momentum}]. When $\mathcal K = 0$ one recovers the conservative motion. 
Moreover, for perturbative systems with small nonconservative coupling $\mathcal K$, one may Taylor‐expand the shifted arguments of $\hamstd$ in \eqref{eq:hamext}, allowing the conservative and nonconservative contributions to be treated independently.

The conservative and nonconservative contributions can also be split by performing an appropriate Taylor expansion around the physical limit, which does not affect the physical trajectories (as shall be seen in the next section). 
We introduce an alternate conservative momentum definition $\tilde {\vec p}_a$, depending only on the $(\vec q_a, \vec \pi_a)$ half of the phase space 
\begin{equation}
    \tilde {\vec p}_a = \vec \pi_a - \left( \frac{\partial \mathcal K}{\partial \dot{\vec q}_-} \right)_{\!\PL,\, a} \text,
\end{equation}
with the subscript ``$\PL, a$'' indicating that the expression is to be evaluated in the physical limit and subsequently replaced with labels $a \in \{\labelA, \labelB\}$. 
It is simple to see that under the physical limit, these momenta reduce to $\tilde {\vec p}_a~=~\vec p$. Then, performing a Taylor-expansion of $\mathcal H$ around $\pathAB{\tilde{\vec p}}$, we extract the more manageable form for $\mathcal A$: 
\begin{align}
    \hamext(\pathany{\vec q}, \pathany{\vec \pi}, t) & = 
    \hamstd (\pathA{\vec q}, \pathA{\tilde{\vec p}}, t )
    - \hamstd (\pathB{\vec q}, \pathB{\tilde{\vec p}}, t )\nonumber\\
    & + \tilde{\mathcal R}(\pathany{\vec q}, \pathany{\vec \pi}, t) \text,\label{eq:hamext_alt}
\end{align}
where we have introduced $\tilde{\mathcal R}$, which equals
\begin{align}
    \tilde{\mathcal R}(\pathany{\vec q}, \pathany{\vec \pi}, t) & = \eta_{ab} \, \dot{\vec q}_a \cdot \left( \dext{\mathcal K}{\dot{\vec q}_-} \right)_{\!\PL\,, b\!\!\!} \nonumber\\
    & - \,\, \mathcal K(\pathany{\vec q}, \pathany{\vec \pi}, t) \text.
\end{align}
Because $\tilde{\vec p}_a$ depends only on $(\vec q_a, \vec \pi_a)$, in this form, the conservative sector is manifestly separable into two
independent copies of $\hamstd$, while all nonconservative corrections are collected into $\tilde{\mathcal R}$. In practical applications, it presents a more tractable form in which to perform computations rather than Eq.\ \eqref{eq:hamext}.

With this we have provided corrections to the nonconservative Hamiltonian and equations of motion computed by Galley \cite[Eq.\ (12) and (17)]{Galley2013}, which we are unable to reproduce. In the following sections, we develop our Hamiltonian framework and demonstrate its many properties.
In Sec.~\ref{sec:canonical_gauge_shift}, we shall show that our nonconservative Hamiltonian is only unique up to a canonical gauge shift. We shall further identify a special form of the Legendre transformation, associated to a specific gauge, which is described expressly in terms of the conservative momentum $\vec p_a$.

\subsection{Symplectic structure}

In order to develop the Hamiltonian formulation for nonconservative systems, it is useful to explicitly construct an appropriate symplectic structure on the extended phase space. Recall that our doubled phase space comprises two copies of the original degrees of freedom, with canonical pairs $\pathA{\vec x} = \tr{(\pathA{\vec q}, \pathA{\vec \pi})}$ and $\pathB{\vec x} = \tr{(\pathB{\vec q}, \pathB{\vec \pi})}$. 
The canonical Hamiltonian equations [Eqs.\ \eqref{eq:ham_eqs}] can be re-expressed as
\begin{equation}
  \dot{\vec x}_a = \vec X_a^{\hamext} (\vec x_b) \, , \label{eq:ham_eqs_X}
\end{equation}
where we have just introduced the Hamiltonian vector field $\vec X_a^{\hamext}$, given by
\begin{equation}
   \vec X_a^{\hamext} = \pathmetric \, \mat S \, \grad_{\!b} \, \hamext \, .
   \label{eq:Hamiltonian_flow}
\end{equation}
In the last expression, $\mat S$ is the symplectic matrix and $\grad_{\!b}$ is the phase-space gradient with respect to $\vec x_b$'s components; they are respectively given by $\mat S = \antidiag(\idmat{N},-\idmat{N})$ and $\grad_{\!b} = \tr{(\partial/\partial \vec q_b,\partial/\partial \vec \pi_b)}$.

Accordingly, the extended Poisson bracket between any two scalars $f$ and $g$ on the doubled phase space becomes
\begin{align}
  \lPB f, g \rPB
  &= \pathmetric \, \tr{\left(\grad_{\!a} f\right)} \, \mathbf S \, \left(\grad_{\!b} \, g\right) \nonumber\\
  &= \pathmetric \left( \frac{\partial f}{\partial \vec q_a} \cdot \frac{\partial g}{\partial \vec \pi_b} - \frac{\partial f}{\partial \vec \pi_a} \cdot \frac{\partial g}{\partial \vec q_b} \right) \text.
    \label{eq:poisson_bracket}
\end{align}
Despite the unconventional presence of the metric $\pathmetric$, the bracket \eqref{eq:poisson_bracket} is fully symplectic, satisfying bilinearity, anti-symmetry, and the Jacobi and Leibniz identities. A useful standard consequence is the derivation property:
\begin{equation}
    \lPB f , g \circ \vec h \rPB = \lPB f, \vec h \rPB \cdot \frac{\partial}{\partial \vec h} (g \circ \vec h) \text.
\end{equation}

In the Hamiltonian framework one can also freely pass between the $(\labelA, \labelB)$ and $(+, -)$ parametrisation [Eqs. (\ref{eq:pm_variables_q}, \ref{eq:pm_variables_p})]. 
However, since the Poisson bracket structure gets modified, the variable transformation $(\labelA, \labelB) \to (+, -)$ is by definition not a canonical transformation.
Namely, in either coordinate system, the Poisson algebra can be described simply via:
\begin{equation}
    \lPB \vec q_a, \vec \pi_b \rPB = \pathmetric \, \idmat{\!N}.
\end{equation}
It is enlightening to explicitly view this Poisson algebra in each variable set. In $(\labelA, \labelB)$ variables we recover
\begin{subequations}
\begin{align}
    &\lPB \pathA{\vec q}, \pathA{\vec \pi} \rPB = \idmat{\!N}, &
    &\lPB \pathA{\vec q}, \pathB{\vec \pi} \rPB = 0, \\
    &\lPB \pathB{\vec q}, \pathB{\vec \pi} \rPB = - \idmat{\!N}, &
    &\lPB \pathB{\vec q}, \pathA{\vec \pi} \rPB = 0.
\end{align}
\end{subequations}
The negative sign conspires with the label-exchange anti-symmetry of the Hamiltonian to ensure that both paths $(\pathA{\vec q}, \pathA{\vec \pi})$ and $(\pathB{\vec q}, \pathB{\vec \pi})$ evolve with the same dynamics.
When transitioning to $(+,-)$, we instead recover a canonical structure where the conjugate pairs are $(\vec q_+, \vec \pi_-)$ and $(\vec q_-, \vec \pi_+)$:
\begin{subequations}
\begin{align}
    &\lPB \vec q_+, \vec \pi_- \rPB = \idmat{N}, &
    &\lPB \vec q_+, \vec \pi_+ \rPB = 0, \\
    &\lPB \vec q_-, \vec \pi_+ \rPB = \idmat{N}, &
    &\lPB \vec q_-, \vec \pi_- \rPB = 0.
\end{align}
\end{subequations}

The dynamical equations of motion follow. Indeed, if we denote the double Hamiltonian by $\hamext$, then Hamilton’s equations for any phase-space function can be extracted via
\begin{equation}
    \frac{\dd f}{\dd t} = \frac{\partial f}{\partial t} + \lPB f , \hamext \rPB \text.\label{eq:poisson_evolution}
\end{equation}

In the Hamiltonian framework, analogous arguments to those provided in Sec.~\ref{sec:nonconservative-lagrangian-formulation} ensure that the physical solution arises as the unique solution from the symmetry properties of the double Hamiltonian. Namely, in the $(\labelA, \labelB)$ variables, the Hamiltonian $\hamext$ is antisymmetric under the $\pathinterchange$ label exchange; accordingly, in the $(+, -)$ variable set, $\hamext$ contains only odd order elements in the expansion in $(-)$ variables. An interesting straightforward corolary is that the value of Hamiltonian $\hamext$ (Noether energy) is identically zero in the physical trajectories.

For coherence, we may redefine the physical limit manifold as the slice of phase space solutions
\begin{multline}
\Sigma=\big\{  (\vec q_a, \vec \pi_a) \in \mathcal C_2([\ts, \te], \mathbb{R}^{4N}) ; \\
\pathA{\vec q} = \pathB{\vec q} \equiv \vec q, \pathA{\vec \pi} = \pathB{\vec \pi} \equiv \vec \pi \big\} \, .    
\end{multline}
Accordingly, the physical limit is now expressed by the conditions $(\vec q_+, \vec \pi_+) = (\vec q, \vec \pi)$ and $(\vec q_-, \vec \pi_-) = (0, 0)$.
As in the Lagrangian framework, it follows that only terms in the Hamiltonian that are linear in the minus variables contribute to the motion and physical observables. 
Any terms of order $\vorder{2}$ will vanish when the physical limit is imposed. This observation justifies a perturbative expansion of the Hamiltonian in powers of the minus variables and is a powerful tool for simplifying the analysis of nonconservative systems.

\subsubsection{Additional Notation and Perturbative Expansions} \label{sec:hamiltonian_properties}

To complete our discussion of the symplectic framework on the doubled phase space, it proves convenient to introduce a set of notational conventions that make perturbative expansions in the minus variables transparent. This notation helps isolate genuine physical contributions of the dynamics---those surviving in the physical limit---from redundant ``virtual'' terms.

We begin by recalling the standard Poisson bracket on the single (physical) phase-space. For two smooth functions $f = f(\vec q, \vec \pi, t)$ and $g = g(\vec q, \vec \pi, t)$, we follow the convention
\begin{align}
  \{ f, g \} = \frac{\partial f}{\partial \vec q} \cdot \frac{\partial g}{\partial \vec \pi} - \frac{\partial f}{\partial \vec \pi} \cdot \frac{\partial g}{\partial \vec q} \text.
\end{align}
Next, we employ the labels $(\labelA, \labelB)$ to denote their evaluation in each section of the doubled phase space
\begin{align}
    \pathA{f} = f(\pathA{\vec q}, \pathA{\vec \pi}, t) \text, && \pathB{f} = f(\pathB{\vec q}, \pathB{\vec \pi}, t) \text,
\end{align}
and form the symmetric and antisymmetric $(+, -)$ combinations,
\begin{align}
    f_+ = \frac{1}{2}(\pathA{f} + \pathB{f}) \text, && & f_- = \pathA{f} - \pathB{f} \text.
\end{align}
A straightforward calculation shows that the doubled‐space bracket obeys
\begin{subequations}\label{eq:pm_vars_bracket_algebra}
\begin{align}
    \lPB f_+, g_- \rPB & = \{ f, g \}_+  \text, \\
    \lPB f_-, g_- \rPB & = \{ f, g \}_- \text, \\
    \lPB f_+, g_+ \rPB & = \{ f, g \}_- / 4 \text,
\end{align}
\end{subequations}
where $\{f,g\}_{\pm}$ indicates the corresponding $\pm$ combination applied to the single‐space bracket.

To see how physical and virtual orders arise, we perform a Taylor expansion about the physical limit ($\vec q_- = 0$ and $\vec \pi_- = 0$). Denoting the single phase space coordinates collectively as $\vec x = (\vec q, \vec \pi)$, we obtain
\begin{subequations}\label{eq:f_pm_properties}
\begin{align}
    & f_+ = f(\vec x_+) + \vorder{2} \text, \\
    & f_- = \vec x_- \cdot \frac{\partial f}{\partial \vec x}(\vec x_+) + \vorder{3} \text,
\end{align}
\end{subequations}
where $\vec x_\pm$ are the different plus-minus combinations.
More generally, $\vec x$ can be an arbitrary set of independent variables forming a diffeomorphism in phase-space, that is, $\vec x = \vec x(\vec q, \vec \pi)$. In this case, the chain rule yields an analogous form:
\begin{equation}
    f_- = \vec x_- \cdot \left( 
    \frac{\partial \vec q}{\partial \vec x} \, \frac{\partial f}{\partial \vec q} 
    + \frac{\partial \vec \pi}{\partial \vec x} \, \frac{\partial f}{\partial \vec \pi}
    \right)_{\vec x_+\!\!\!} + \vorder{3} \text, \label{eq:-_expansion}
\end{equation}
where the terms within the parenthesis are evaluated at $\vec x = \vec x_+$.
These expansions underpin the identification of physical (order-one) versus virtual (higher-order) contributions in the doubled‐variable formalism.

More generally, any function anti-symmetric with respect to $(\pathA{\vec q}, \pathA{\vec \pi}) \leftrightarrow (\pathB{\vec q}, \pathB{\vec \pi})$ label exchange has odd order in $\vec x_-$ and can be expanded similarly to $f_-$. 
Analogously, any symmetric function has even order in $\vec x_-$ and can be expanded similarly to $f_+$.
Since the nonconservative Hamiltonian $\hamext$ is by construction odd under copy exchange, its linearisation about the physical slice takes the form
\begin{equation}
    \hamext(\vec q_\pm, \vec \pi_\pm) = \vec x_- \cdot \hamext_{\vec x}(\vec x_+) + \vorder{3} \text, \label{eq:hamiltonian_perturbative_eom_form}
\end{equation}
where the Hamiltonian coefficients $\hamext_{\vec x} = (\hamext_{\vec q}, \hamext_{\vec \pi})$ are newly introduced functions of $(\vec q_+, \vec \pi_+)$. They can be identified from Hamilton's equations of motion [Eq.\ \eqref{eq:poisson_evolution}] applied to $a = +$ in the physical limit:
\begin{subequations}
\begin{empheq}[left=\empheqlbrace]{align}
     \hamext_{\vec q}(\vec q, \vec \pi) & = \! \left( \,\dext{\hamext}{\vec q_-} \right)_{\PL\!} \!\! = - \dot{\vec \pi}(\vec q, \vec \pi) \text, \\
     \hamext_{\vec \pi}(\vec q, \vec \pi) & = \! \left( \,\dext{\hamext}{\vec \pi_-} \right)_{\PL\!} \!\! = \dot{\vec q}(\vec q, \vec \pi) \text,
\end{empheq}
\end{subequations}
so that the known kinematics and generalized forces of a given system immediately determine the linear structure of $\hamext$.  We shall exploit this strategy for constructing $\hamext$ in applications (see Sec.~\ref{sec:inverse-variational-problem}).

Finally, any extended Hamiltonian $\mathcal A$ may be split into a conservative component $\tilde{\mathcal H}_-$ and a non-conservative remainder $\tilde{\mathcal R}$, either by Helmholtz decomposition (Appendix \ref{sec:Helmholtz}) or through the alternative form of the Legendre transform [Eq.\ \eqref{eq:hamext_alt}].
On the physical slice, the evolution of an observable $f(\vec x)$ then assumes the succinct form
\begin{equation}
    \frac{\dd f}{\dd t} = \frac{\partial f}{\partial t} + \{ f , \tilde{\mathcal H} \} + \{ f, \vec x \} \cdot \tilde{\mathcal R}_{\vec x} \text,
\end{equation}
cleanly separating the symplectic flow generated by $\tilde{\mathcal H}$ from the non-conservative contributions encoded in the components 
$$\tilde{\mathcal R}_{\vec x} = \left(\,\frac{\partial \tilde{\mathcal R}}{\partial x_- \!\!}\,\right)_{\!\PL} \text. $$

\subsubsection{Liouville volume}

One of the most remarkable implications of the doubled-variable Hamiltonian formulation is 
its ability to reconcile nonconservative dynamics with the incompressible flows mandated by Liouville’s theorem. 
Naively, this looks contradictory, since these systems must generally lose energy and contract their phase space, whereas Hamiltonian systems preserve both energy (in its canonical sense) and phase‐space volume.

Explaining the non-conservation of energy is straightforward: the value of the doubled Hamiltonian is always zero on the physical trajectories, due to the anti-symmetric construction of the function. Hence, this value is always conserved; it is dissociated from the physical energy of the system.

The second point is more subtle. 
In Hamiltonian mechanics, Liouville’s theorem states that the flow preserves the total phase‐space volume in its usual symplectic measure. In other words, any finite‐volume region will be dragged along by the dynamics without any net expansion or contraction over time.
We know that dissipative systems typically violate Liouville’s theorem: they exhibit compressibility, causing trajectories to converge or spiral into attractors (fixed points, limit cycles, etc.). Hamiltonian flows, on the other hand, are incompressible, and cannot simply collapse all trajectories into a single point.

However, in the double-variable framework, the physical degrees of freedom live on a submanifold of strictly lower dimension: a $2 M$-dimensional subset of a $4 M$-dimensional space. From the higher‐dimensional perspective, this submanifold has measure zero in the ambient space. This embedding allows trajectory curves that preserve volume of the \emph{global space}, yet, when viewed from within the physical slice $\PL$, this cross-section can contract or expand.

This can be illustrated by computing the divergence of the Hamiltonian flow [cf.\ Eq.\ \eqref{eq:Hamiltonian_flow}]. Indeed, we can verify that, in accordance to Liouville's theorem, the doubled Hamiltonian vector field remains divergence-free:
\begin{align}
    \delta_{ab} \, \grad_a \cdot \vec X_b^{\hamext} & = \pathmetric \left( \dext{}{\vec q_a} \cdot \dext{\hamext}{\vec \pi_b} - \dext{}{\vec \pi_a} \cdot \dext{\hamext}{\vec q_b} \right) =0 \, ,
\end{align}
where the last equality stems from the commutative property of the partial derivatives. Invoking Hamilton's equations [Eq.~\eqref{eq:ham_eqs_X}], Liouville's theorem for the doubled-variable system reads as
\begin{align}
    \delta_{ab} \left( \frac{\partial}{\partial \vec q_a} \cdot \dot{\vec q}_b + \frac{\partial}{\partial \vec \pi_a} \cdot \dot{\vec \pi}_b \right) = 0 \, .
\end{align}
In particular, when expressed in $(+,-)$ variables, 
\begin{equation}
   \dext{}{\vec q_+} \cdot \dot{\vec q}_+ + \dext{}{\vec \pi_+} \cdot \dot{\vec \pi}_+ + \dext{}{\vec q_-} \cdot \dot{\vec q}_- + \dext{}{\vec \pi_-} \cdot \dot{\vec \pi}_- = 0 \, .
   \label{eq:div_liouville}
\end{equation}
On the other hand, 
we see that the divergence of the vector field restricted to the physical slice $\Sigma$ can in general be non-zero:
\begin{equation}
   \grad_+ \cdot \vec X_+^{\hamext} = \dext{}{\vec q_+} \cdot \dot{\vec q}_+ + \dext{}{\vec \pi_+} \cdot \dot{\vec \pi}_+ \, ,
\end{equation}
since it only contains part of the terms of the total divergence in Eq.\ \eqref{eq:div_liouville}.
The physical limit solution implies that the submanifold $\Sigma$ is invariant under the flow: a region contained inside $\Sigma$ will have its evolution constrained within $\Sigma$ at all future times. Hence, trajectories initiating in the physical slice $\Sigma$ will exhibit genuinely nonconservative dynamics. 

\subsubsection{Example:}

\begin{figure*}
    \centering
    \includegraphics[scale=0.52]{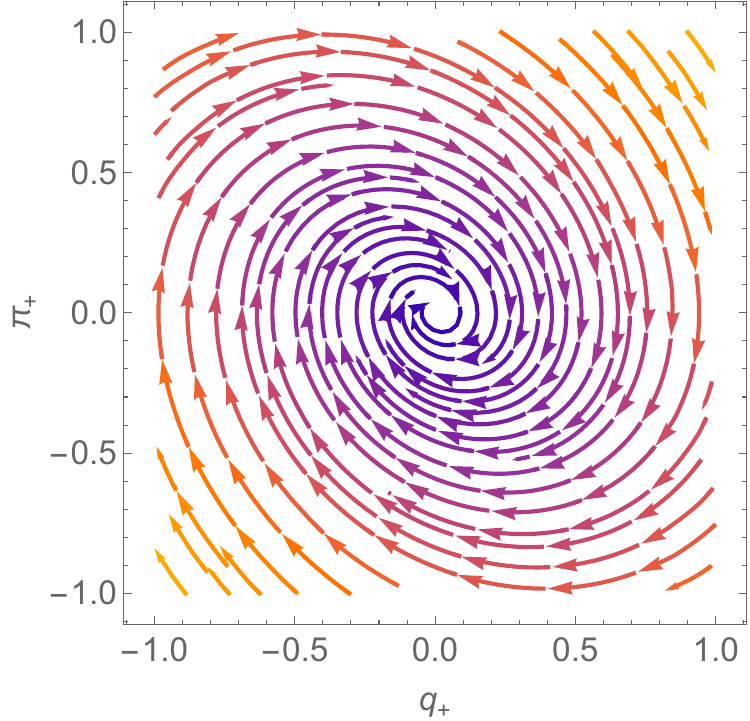}
    \hspace{0.5cm}
    \includegraphics[scale=0.52]{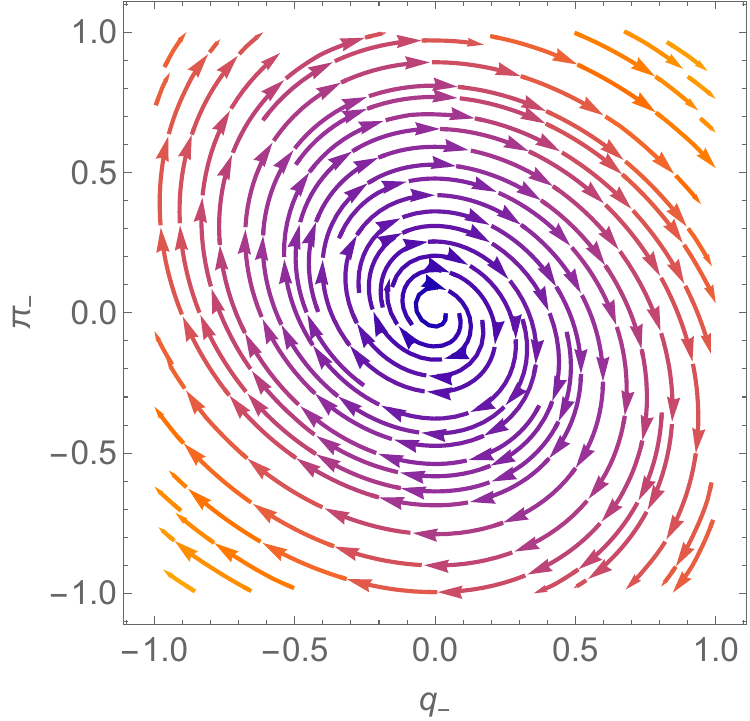}
    \caption{Phase-space slices for the damped harmonic oscillator in the doubled-variable Hamiltonian formalism.
    On the \emph{right}: the physical slice $(q_+,\pi_+)$, a submanifold invariant to the dynamical flow, where the virtual trajectories are constrined to $(q_-,\pi_-) = (0,0)$. Dissipation causes all physical trajectories to spiral into the origin, making it a global \emph{attractor}.
    On the \emph{left}: virtual planes $(q_-,\pi_-)$ for fixed $(q_+,\pi_+)$; the flow diverges from the origin, which here acts as a \emph{repeller}.
    The color map represents the vector field magnitude $\norm{(\dot q_\pm, \dot \pi_\pm)}$: warm tones (orange or light gray) mark large velocities, approaching cool tones (blue or dark gray) for vanishing velocities near the origin---which is a fixed point.}
    \label{fig:harmonic_oscillator_phase_space} 
\end{figure*}

As an illustrative example, we consider the damped harmonic oscillator, which can be described by the Hamiltonian (see Sec.~\ref{sec:inverse-variational-problem} for details on how to construct it from the equations of motion):

\begin{equation}
    \hamext(q_\pm, \pi_\pm) = \frac{1}{m} \pi_- \pi_+ + q_- \left(\frac{\gamma}{m} \pi_+ + k q_+ \right) \text,
\end{equation}
and the following equations of motion:
\begin{subequations}
\begin{align}
    & \dot q_+ = \frac{\pi_+\!}{m} \text, &
    & \dot q_- = \frac{\pi_-\!}{m} + \frac{\gamma}{m} q_- \text, \\ 
    & \dot \pi_+ = - \frac{\gamma}{m} \pi_+ - k q_+ \text, &
    & \dot \pi_- = - k q_- \text.
\end{align}
\end{subequations}
The origin is a fixed point, where the Hamiltonian vector field is zero.
As can be seen in Fig.~\ref{fig:harmonic_oscillator_phase_space}, the physical slice contains an attractor at the origin, which shrinks the phase space cross-section. Meanwhile, the orthogonal subspace to $\Sigma$ contains a repeller. Accordingly, the total divergence of the system evaluates to zero, whereas the divergence of the physical slice reduces to
\begin{equation}
    \grad_+ \cdot \vec X_+^{\hamext} = \dext{\dot q_+\!}{q_+} + \dext{\dot \pi_+\!}{\pi_+} = -\frac{\gamma}{m} \, .
\end{equation}


\subsection{Linearized Hamiltonian}\label{sec:linear_Ham}

Replacing the Lagrangian in Lie form [see Eq.~\eqref{eq:linearized_action_Lie_form}] into the Legendre transform \eqref{eq:nonconservative_Legendre_transform} yields a linearized Hamiltonian
\begin{equation}
    \hamext(\vec q_\pm, \vec \pi_\pm, t) = \vec \pi_- \cdot \dot{\vec q}_+ - \vec q_- \cdot \Dt[{\vec \pi}_+] \text, \label{eq:linear_hamext} 
\end{equation}
where $\dot{\vec q}_+$ and $\Dt[{\vec \pi}_+]$ must now be expressed in terms of phase-space variables. Namely, we define at $\vorder{2}$
\begin{subequations}
\begin{align}
    \Dt[\vec q_+] &= \, \dot{\vec q}_+ \,\,\, \equiv \vec V(\vec q_+, \vec \pi_+, t) \text, \\
    \Dt[\vec \pi_+] &= \dext{\lagext}{\vec q_-} \equiv \vec G(\vec q_+, \vec \pi_+, t) \text,
\end{align}
\end{subequations}
where $\vec V$ and $\vec G$ are obtained via the usual Legendre inversion of the physical momentum expression. In this way, the Hamiltonian \eqref{eq:linear_hamext} becomes
\begin{equation}
    \hamext(\vec q_\pm, \vec \pi_\pm, t) = \vec \pi_- \cdot \vec V(\vec q_+, \vec \pi_+, t) - \vec q_- \cdot \vec G(\vec q_+, \vec \pi_+, t) \text. \label{eq:linear_hamext_VG} 
\end{equation}

For completeness, we also include the on-shell derivative for the $(-)$ phase-space variables, which are similarly computed from inversion of Eqs.~\eqref{eq:nonconservative_momentum} and \eqref{eq:onshell_dp_dt}, with the Lagrangian in linear form [Eq.~\eqref{eq:linearized_action_Lie_form}]:
\begin{subequations}
\begin{align}
    \Dt[\vec q_-] &= \,\vec\pi_- \cdot \dext{\vec V}{\vec \pi_+} - \vec q_- \cdot \dext{\vec G}{\vec \pi_+} \text, \\
    \Dt[\vec \pi_-] &= - \vec\pi_- \cdot \dext{\vec V}{\vec q_+} + \vec q_- \cdot \dext{\vec G}{\vec q_+}  \text.
\end{align}
\end{subequations}
More generally, these on-shell derivatives directly connect to the partial derivatives of $\hamext$ and hence to the Hamiltonian equations:
\begin{subequations}\label{eq:onshell_dt_phasespace_coeff}
\begin{align}
    \Dt[\vec q_a] &= \dot{\vec q}_a = \pathmetric[ab] \dext{\hamext}{\vec \pi_b} \text, \\
    \Dt[\vec \pi_a] &= \pathmetric[ab] \dext{\lagext}{\vec q_b} = - \pathmetric[ab] \dext{\hamext}{\vec q_b} \text.
\end{align}
\end{subequations}

We are now ready to identify $\Dt$ as a total time derivative along the phase-space dynamics. For this, since $\Dt$ must satisfy the derivation property, we consider in phase-space
\begin{equation}
    \Dt = \frac{\partial}{\partial t} + \delta_{ab} \left( \Dt[\vec q_a] \cdot \dext{}{\vec q_b} + \Dt[\vec \pi_a] \cdot \dext{}{\vec \pi_b} \right) \text. \label{eq:onshell_dt_phasespace_with_coeff}
\end{equation}
Plugging the coefficients $\Dt[\argdot]$ from Eq.~\eqref{eq:onshell_dt_phasespace_coeff}, we recover the Lie-Poisson derivative along the flow of $\hamext$ and time
\begin{equation}
    \Dt = \frac{\partial}{\partial t} + \lPB\argdot, \hamext\rPB \text.\label{eq:onshell_dt_phasespace}
\end{equation}

The linearized forms of $\lagext$ [via Eq.~\eqref{eq:linearized_action_Lie_form}] and of $\hamext$ [Eq.\ \eqref{eq:linear_hamext}] allow a trivial reconstruction from a given set of (not necessary symplectic) ODEs. This can be achieved by prescribing a momentum definition $\vec \pi_+ = \vec \pi(\vec q_+, \dot{\vec q}_+)$ (or equivalently, the coordinate velocity $\vec V = \vec V(\vec q_+, \vec \pi_+)$), and subsequently identifying the on-shell derivative which describes the system dynamics via the acceleration term $\vec U_+ = \vec U(\vec q, \dot{\vec q}, t)$ or the force $\vec G=\vec G(\vec q, \vec \pi, t)$. More details are available in Sec.\ \ref{sec:inverse-variational-problem}.


\subsection{Canonical gauge shift}
\label{sec:canonical_gauge_shift}

A notable feature of the doubled–variable structure is the freedom to modify the extended Lagrangian by adding terms that do not contribute to the physical trajectories. Although such changes do not affect the observable dynamics, they can modify the canonical quantities derived from the Lagrangian.
In this section, we present an alternative route to the Legendre transformation that involves adjusting how we define the nonconservative momenta. Such adjustments thus act as a kind of ``gauge freedom'' in selecting the nonconservative terms and associated canonical quantities.

It is most transparent to work in the $(+, -)$ variables, discarding all terms of order $\vorder{2}$ since they neither influence the canonical structure nor the observable trajectories. Without loss of generality, one may always revert to the $(\labelA, \labelB)$ variables by inverting Eqs.\ (\ref{eq:pm_variables_q}) and (\ref{eq:pm_variables_p}).

Starting from the nonconservative Lagrangian, we consider a general transformation of the form
\begin{equation}
    \lagext \to \lagext' = \lagext + \vec q_- \cdot \vec \psi + \dot{\vec q}_- \cdot \vec \phi + \vorder{3}.
\end{equation}
Initially, $\vec \psi = \vec \psi (\vec q_+, \dot{\vec q}_+, t)$ and $\vec \phi = \vec \phi (\vec q_+, \dot{\vec q}_+, t)$ are arbitrary vector–valued functions that depend only on the physical variables and time. Different choices of the gauge functions $(\vec \phi, \vec \psi)$ correspond to shifts in the Lagrangian $\lagext \to \lagext'$ and can be interpreted as shifts in the nonconservative term $\mathcal K \to \mathcal K'$.

We recall that in the $(+,-)$ parametrization, the physical path is determined by varying the action with respect to $\vec q_-$, so that the Euler–Lagrange equation for $\lagext'$ picks up an extra term:
\begin{align}
    0 &= \left( \dext{\lagext'}{\vec q_-} - \frac{\dd}{\dd t} \dext{\lagext'}{\dot{\vec q}_-} \right) \nonumber\\
    &= \left( \dext{\lagext}{\vec q_-} - \frac{\dd}{\dd t} \dext{\lagext}{\dot{\vec q}_-} \right) + \big( \vec \psi - \dot{\vec \phi} \big) + \vorder{2} \text. \label{eq:EL_gauge_plus}
\end{align}
To guarantee invariance of the physical dynamics under the gauge shift, one must demand that the extra term on the RHS of Eq.~\eqref{eq:EL_gauge_plus} vanishes on-shell.
The naive requirement 
\begin{equation}
    \vec \psi = \frac{\dd \vec \phi}{\dd t} = \frac{\partial \vec \phi}{\partial t} + \dot{\vec q}_+ \cdot \dext{\vec \phi}{\vec q_+} + \ddot{\vec q}_+ \cdot \dext{\vec \phi}{\dot{\vec q}_+} \text.  \label{eq:gauge_psi_naive}
\end{equation}
introduces accelerations into the Lagrangian, which we prefer to avoid\footnote{We remark in passing that in the double-variable formulation the situation is more favourable than for standard action principles. Specifically, the introduction of plus-variable accelerations $\ddot{\vec q}_+$ into the nonconservative Lagrangian $\lagext$ though a piece of the form $\vec q_- \cdot \vec f(\vec q_+, \dot{\vec q}_+, \ddot{\vec q}_+, t)$ still leads to Euler-Lagrange equations of second order, and hence averts the risks of Ostrogradsky instabilities. This can be verified directly from the physical Euler-Lagrangian equations [Eq.~\eqref{eq:EL_nonconservative_physical}]. Moreover, there is not even any need for modifications of the boundary conditions of the action principle. Namely, the initial conditions for $\vec q_+$ and $\dot{\vec q}_+$ still provide complete Cauchy data for the second order physical Euler-Lagrange equation; meanwhile, the equality conditions defined via $\vec q_-$ still lead to vanishing variation boundaries and the physical limit.}. 
The remedy is to replace the ordinary time derivative in Eq.~\eqref{eq:gauge_psi_naive} by the on–shell derivative $\Dt$ [see Eq.~\eqref{eq:onshell_dt}], where accelerations are replaced by their dynamical expressions in terms of $\vec q_a$ and $\dot{\vec q}_a$.
We then impose:
\begin{equation}
    \vec \psi = \Dt [\vec \phi] 
    = \frac{\dd \vec \phi}{\dd t} + \mat H^{-1}_{-+} \left( \dext{\lagext}{\vec q_-} - \frac{\dd}{\dd t} \dext{\lagext}{\dot{\vec q}_-} \right) \cdot \dext{\vec \phi}{\dot{\vec q}_+}
    \text, \label{eq:gauge_psi}
\end{equation}
where on the RHS we have employed the explicit expression of $\vec U_+$ [Eq.~\eqref{eq:on_shell_dt_G}] after comparing the operators $\Dt$ and $\dd/\dd t$ in expanded form.

This choice of $\vec \psi$ ensures that the physical Euler–Lagrange equation for the shifted Lagrangian $\lagext'$ [Eq.~\eqref{eq:EL_gauge_plus}] and the unshifted Lagrangian $\lagext$ [Eq.~\eqref{eq:EL_nonconservative_physical}] are equivalent.
Namely, if the equation for $\lagext$ is satisfied, then $\Dt[\phi] - \dot{\vec \phi}$ vanishes and Eq.~\eqref{eq:EL_gauge_plus} is satisfied. Conversely, if the physical Euler–Lagrange equation for $\lagext'$ holds, then the one for $\lagext$ follows, as we shall demonstrate. Specifically, using Eq.~\eqref{eq:gauge_psi}, the Euler–Lagrange equation for $\lagext'$ may be written in the form:
\begin{equation}
    \mat{R_\phi} \, \left( \dext{\lagext}{\vec q_-} - \frac{\dd}{\dd t} \dext{\lagext}{\dot{\vec q}_-} \right) = 0 \text,
\end{equation}
with
\begin{equation*}
    \mat{R_\phi} = \left( \mathbbm{1}_N + \tr{\!\left( \, \dext{\vec \phi}{\dot{\vec q}_-} \right)} \, \mat H_{-+}^{-1} \right) \text.
\end{equation*}
We can identify the product $\mat{R_\phi} \mat H_{-+}$ with the Hessian of $\lagext'$. Since both $\lagext$ and the gauge–transformed Lagrangian $\lagext'$ are assumed regular, it follows that $\mat{R_\phi}$ is invertible. Thus, the Euler–Lagrange equations for $\lagext'$ imply those for $\lagext$: any extra terms from the gauge change can be ``divided out'', so that the equations of motion are indeed equivalent.

Although the motion itself remains unaffected, the canonical momenta do not. To first order in $(-)$, one finds:
\begin{subequations}
\begin{align}
    &\vec \pi_+' \equiv \dext{\lagext'}{\dot{\vec q}_-} = \vec \pi_+ + \vec \phi + \vorder{2} \text, \label{eq:pi_plus_gauge} \\
    &\vec \pi_-' \equiv \dext{\lagext'}{\dot{\vec q}_+} = \vec \pi_- + \dext{\vec \psi}{\dot{\vec q}_+} \cdot \vec q_- + \dext{\vec \phi}{\dot{\vec q}_+} \cdot \dot{\vec q}_- + \vorder{3} \text, \label{eq:pi_minus_gauge}
\end{align}
\end{subequations}
so that the physical momentum $\vec \pi_+$ is simply shifted by the gauge function $\vec \phi$, while the virtual momentum $\vec \pi_-$ picks up first–order corrections---in agreement with the computations in Sec.\ \ref{sec:linear:gauge_shift} [cf.\ Eq.\ \eqref{eq:pi_gauge}].
Although seemingly innocuous, these terms will impact the gauge transformed Hamiltonian. A direct Legendre transform for $\Lambda'$ shows that
\begin{align}
    \hamext'(\vec q_\pm, \vec \pi_\pm', t) & = \hamext(\vec q_\pm, \vec \pi_\pm, t) 
    + \left( \dot{\vec q}_+ \cdot \dext{{\vec \psi}}{\dot{\vec q}_+} - \vec \psi \right) \cdot \vec q_-\nonumber\\
    &+ \left( \dot{\vec q}_+ \cdot \dext{\vec \phi}{\dot{\vec q}_+} \right) \cdot \dot{\vec q}_- + \vorder{3} \text. \label{eq:hamext_gauge_v0}
\end{align}
Importantly, at this stage the RHS contains dependencies with respect to the unshifted momentum variables inside $\hamext(\vec q_\pm, \vec \pi_\pm, t)$. To express the new Hamiltonian entirely in terms of the shifted phase–space variables, we consider a first–order Taylor expansion of $\hamext(\vec q_\pm, \vec \pi_\pm, t)$ with respect to the virtual momentum $\vec \pi_-$ around the shifted value $\vec \pi_-'$:
\begin{equation}
    \hamext(\vec q_\pm, \vec \pi_\pm, t) = \hamext(\vec q_\pm, \vec \pi_+, \vec \pi_-', t) + (\vec \pi_- - \vec \pi_-') \cdot \dot{\vec q}_+  + \vorder{3} \text,
\end{equation}
where have replaced $\partial \hamext / \partial \vec \pi_- = \dot{\vec q}_+$ at leading order.
After substitution [together with Eq.\ \eqref{eq:pi_minus_gauge}], the gauge-transformed Hamiltonian \eqref{eq:hamext_gauge_v0} collapses into the compact form:
\begin{equation}
    \hamext'(\vec q_\pm, \vec \pi_\pm', t) = \hamext(\vec q_\pm, \vec \pi_+' - \vec \phi, \vec \pi_-', t) - \vec q_- \cdot \vec{\psi}  + \vorder{3} \text. \label{eq:hamext_gauge}
\end{equation}
The gauge terms $\vec{\phi}$ and $\vec{\psi}$ must, of course, also be expressed in terms of the (shifted) phase space variables $(\vec q_+, \vec \pi_+')$. As usual, this is achieved by inverting the momentum relation---now Eq.~\eqref{eq:pi_plus_gauge}---with respect to $\dot{\vec q}_+$ and substituting into the gauge term expressions. 
Alternatively, once $\vec \phi$ is reframed in terms of the phase space variables, $\vec \psi$ can be obtained from the Poisson form of the on-shell derivative [cf.\ Eq.\ \eqref{eq:onshell_dt_phasespace}].

In this new Hamiltonian, the structure of Hamilton’s equations in the physical limit simply reflects the momentum shift:
\begin{subequations}\label{eq:ham_eqs_gauge}
\begin{align}
    &\dot{\vec q} = \left(\dext{\hamext}{\vec \pi_-}\right)_{\!\PL} \text,  \\
    &\dot{\vec \pi}' = - \left(\dext{\hamext}{\vec q_-}\right)_{\!\PL\!}\! + \Dt[\vec \phi] \text,
\end{align}
\end{subequations}
where in the RHS the Hamiltonian $\mathcal A$ must be evaluated at $\vec \pi = \vec \pi' - \vec \phi$.
Clearly, Eqs.~\eqref{eq:ham_eqs_gauge} are completely equivalent to the unshifted equations [Eqs.~\eqref{eq:ham_eqs}]: the correction $\Dt[\vec\phi]$ in Hamilton's second equation emerges as a direct consequence of the momentum shift by $\vec\phi$.
In conservative mechanics such freedom to redefine momenta while keeping the configuration variables $\vec q$ intact is generally forbidden by the symplectic structure; here, however, the doubled formulation automatically preserves the canonical two‐form by introducing corrections into the Hamiltonian which are reflected in the canonical equations to the virtual $(-)$ variables.

Thus, any convenient choice of $\vec \phi = \vec \phi(\vec q_+,\dot{\vec q}_+, t)$ yields a family of physically equivalent Hamiltonians, differing only by gauge terms in the nonconservative sector. In practice, one prescribes $\vec \phi$ to simplify either the momentum definitions or the Hamiltonian itself, then deduces $\vec \psi=\Dt[\vec \phi]$ by computing the Lie derivative. Since $\vec \phi$ is unconstrained, the physical momentum can be arbitrarily modified via Eq.~\eqref{eq:pi_plus_gauge}. We will show in Sec.~\ref{sec:inverse-variational-problem} how to exploit this freedom in order to reconstruct a nonconservative Hamiltonian directly from given equations of motion.

The form of the shifted Hamiltonian presented in Eq.\ \eqref{eq:hamext_gauge} is useful when working in the $(+,-)$ parametrization, since the gauge shift $\vec \phi$ was defined as a function of $(+)$ variables explicitly. When working in the $(\labelA, \labelB)$ variables---for instance, with a Hamiltonian of the form \eqref{eq:hamext_alt}---it can be more practical to instead expand the virtual momentum in Eq.\ \eqref{eq:hamext_gauge} around $\vec \pi_-' - \vec \phi_-$ (with $\vec \phi_- = \pathA{\vec \phi} - \pathB{\vec \phi}$), upon which we obtain
\begin{multline}
	\hamext'(\pathAB{\vec q}, \pathAB{\vec \pi}', t) = \hamext(\pathAB{\vec q}, \pathAB{\vec \pi}' \mp \pathAB{\vec \phi}, t) 
     \\ + \vec \phi_- \cdot \dot{\vec q}_+ 
      - \vec q_- \cdot \vec{\psi} + \vorder{3} \text. 
\end{multline}
This form allows the independent shifting of the momentum of each path in the arguments of a preexisting Hamiltonian.
The gauge corrections $\vec \phi_- \cdot \dot{\vec q}_+ - \vec q_- \cdot \Dt[\vec \phi]$ take the same form as a linearized Hamiltonian [Eq.~\eqref{eq:linear_hamext}], which is compatible with the remark that the dynamics is fully contained within the operator $\Dt$.

\subsubsection*{Conservative gauge}

A notable choice is the conservative gauge $\vec \phi = - \partial \mathcal K / \partial \dot{\vec q}_-$. Inserting into Eq.~\eqref{eq:pi_plus_gauge} restores the physical definition of the conservative momentum:
\begin{equation}
    \vec \pi_+' = \left(\frac{\partial \mathcal L}{\partial \dot{\vec q}}\right)_{\!+} \!\! = \vec p_+ \text. \label{eq:pi_plus_gauge_cons}
\end{equation}
The corresponding Hamiltonian can be obtained by replacing the expressions of $\vec \psi$ [Eq.~\eqref{eq:gauge_psi}], $\hamext$ [Eq.~\eqref{eq:hamext}] and $\vec \pi_{\pm}'$ [Eq.~\eqref{eq:pi_plus_gauge}] in Eq.~\eqref{eq:hamext} or Eq.~\eqref{eq:hamext_gauge}, then expanding $\mathcal K$ to first order in $(-)$:
\begin{align}
    \hamext(\pathany{\vec q}, \pathany{\vec \pi'}, t) & = \hamstd(\pathA{\vec q}, \pathA{\vec \pi'}, t) - \hamstd(\pathB{\vec q}, \pathB{\vec \pi'}, t) \nonumber\\
    & -\vec q_- \cdot \vec{\mathcal F}(\vec q_+, \vec \pi'_+, t) \text, \label{eq:hamext_gauge_cons}
\end{align}
with $\vec{\mathcal F}$ describing the nonconservative forces:
\begin{equation}
     \vec{\mathcal F}(\vec q, \vec \pi', t) = \left( \dext{\mathcal K}{\vec q_-} - \Dt \!\left[ \dext{\mathcal K}{\dot{\vec q}_-} \right] \right)_{\PL} \text.
\end{equation}
In the above equation, $\dot{\vec q}_+$ must be replaced as function of $(\vec q_+, \vec \pi'_+)$ by inversion of Eq.~\eqref{eq:pi_plus_gauge_cons}.
In the physical limit, the equations of motion are:
\begin{subequations}\label{eq:ham_eqs_gauge_cons}
\begin{align}
    \dot{\vec q} &= \frac{\partial \hamstd}{\partial \vec p} \text, &
    \dot{\vec p} &= - \frac{\partial \hamstd}{\partial \vec q} + \vec{\mathcal F} \text.
\end{align}
\end{subequations}
Observe that the equation for $\dot{\vec q}_+$ remains free of dissipative terms, which is compatible with inverting Eq.~\eqref{eq:pi_plus_gauge_cons} with respect to velocity. All nonconservative effects are fully encapsulated in the momentum equation as the additional force term $\vec{\mathcal F}$.

\section{Flux equations}

Observables must be expressed solely in terms of physical quantities (for instance, $(\vec q, \dot{\vec q}, t)$ or $(\vec q, \vec \pi, t)$). Given an observable $\mathcal U = \mathcal U(\vec q, \vec \pi, t)$, it evolves according to:
\begin{equation}
    \frac{\dd \mathcal U}{\dd t} = \frac{\partial \mathcal U}{\partial t} + \dot{\vec q} \cdot \frac{\partial \mathcal U}{\partial \vec q} + \dot{\vec \pi} \cdot \frac{\partial \mathcal U}{\partial \vec \pi} = \frac{\partial \mathcal U}{\partial t} + \lPB\mathcal U, \hamext\rPB_\PL \text,
\end{equation}
where on the RHS, $\mathcal U$ is seen as function of $(\vec q_+, \vec \pi_+, t)$. 

In the Lagrangian framework, Noether's theorem can be generalized for nonconservative systems by expressing the evolution of conservative Noether charges and currents (obtained from the conservative action) in terms of nonconservative variables \cite[see e.g.][]{Galley2014}. For instance, the conservative energy function
\begin{equation}
    E(\vec q, \dot{\vec q}, t) = \dot{\vec q} \cdot \frac{\partial \mathcal L}{\partial \dot{\vec q}}(\vec q, \dot{\vec q}, t) - \mathcal L(\vec q, \dot{\vec q}, t) \text,
\end{equation}
is nothing less than the conservative Hamiltonian evaluated at the conservative momentum:
\begin{equation}
    E(\vec q, \dot{\vec q}, t) = \hamstd(\vec q, \vec p, t) \text.
\end{equation}
Accordingly, Galley expressed the flux-balance equations in the form:
\begin{equation}
    \frac{\dd E}{\dd t} = - \frac{\partial \mathcal L}{\partial t} + \dot{\vec q} \cdot \vec{\mathcal F} \text,\label{eq:flux_balance}
\end{equation}
which represents the energy dissipated by the conservative elements.
When the nonconservative Hamiltonian system is expressed in the conservative gauge [Eq.\ \eqref{eq:hamext_gauge_cons}], the momentum variable $\vec \pi_+'$ coincides with the conservative $\vec p_+$. In this case, the energy flux can be directly obtained from the Poisson relation:
\begin{equation}
    \frac{\dd \hamstd}{\dd t} = \frac{\partial \hamstd}{\partial t} + \lPB\hamstd, \hamext\rPB \text. \label{eq:flux_balance_hamiltonian}
\end{equation}
Observe that Eqs.\ \eqref{eq:flux_balance} and \eqref{eq:flux_balance_hamiltonian} coincide once the conservative gauge equations of motion have been replaced [Eq.\ \eqref{eq:ham_eqs_gauge_cons}].
In all other gauges, $\hamstd$ must be shifted with respect to its second argument $\hamstd = \hamstd(\vec q, \vec \pi - \dext{K}{\vec q_-}, t)$ when computing the above Poisson bracket.



\section{Reconstruction of variational principles from second-order equations of motion} \label{sec:inverse-variational-problem}

In this section, we demonstrate how to construct a Lagrangian or Hamiltonian consistent with a given set of discrete ordinary differential equations (ODEs). As already hinted by \citeauthor{Bateman1931}'s computations \cite{Bateman1931}, these inverse problems become almost trivial in the double-variable framework. 
In particular we focus on systems of second-order, expressible in the form:
\begin{equation}
\ddot{\vec q}(t) = \vec U(\vec q, \dot{\vec q}, t), \label{eq:reconstruct_principle_acceleration}
\end{equation}
where the on-shell acceleration $\vec U$ may include nonconservative contributions.
A straightforward manner is to designate the Lagrangian
\begin{equation}
    \lagext(\vec q_a, \dot{\vec q}_a, \ddot{\vec q}_+, t) = \vec q_- \cdot \left( \ddot{\vec q}_+ - \vec U(\vec q_+, \dot{\vec q}_+, t) \right) \text, \label{eq:Lagrangian_rec_simple}
\end{equation}
which trivially produces the correct equations of motion. This trick can be applied to any second-order ODE, not only those written in the form \eqref{eq:reconstruct_principle_acceleration}. Observe, however, that we have introduced accelerations into the Lagrangian. As previously discussed, although the boundary conditions in Sec.~\ref{sec:nonconservative-lagrangian-formulation} are assigned for first-order Lagrangians [cf.\ Eq.\ \eqref{eq:boundary_conditions_pmv}], they remain sufficient for describing given by Lagrangian \eqref{eq:Lagrangian_rec_simple}, which yields second-order equations of motion. Yet, the presence of accelerations in the Lagrangian might not be desirable for many physical theories, especially since it hinders the Legendre transformation. 
A simple remedy is to perform an integration by parts, giving
\begin{equation}
    \lagext(\vec q_a, \dot{\vec q}_a, t) = \dot{\vec q}_- \cdot \dot{\vec q}_+ + \vec q_- \cdot \vec U(\vec q_+, \dot{\vec q}_+, t) \text, \label{eq:reconstructed_Lagrangian_simple_2}
\end{equation}
which produces the momentum $\vec \pi_+ = \dot{\vec q}_+$ by construction. Yet by virtue of the freedom to shift the momentum, we know that other Lagrangians are possible. We therefore follow a more general approach, which recovers the above as a special case.

The key step is to recast the system into a first-order form by introducing a momentum-like variable. Owing to the inherent ``canonical gauge freedom'', we can freely define the momentum as convenient:
\begin{equation}
\vec \pi(t) = \vec P(\vec q, \dot{\vec q}, t), \label{eq:reconstruct_principle_momentum}
\end{equation}
as long as $\vec P$ is a local diffeomorphism in its second argument and thus invertible in $\dot{\vec q}$:
\begin{equation}
\dot{\vec q}(t) = \vec V(\vec q, \vec \pi, t). \label{eq:reconstruct_principle_velocity}
\end{equation}
For instance, a trivial choice is to set $\vec \pi(t) = \dot{\vec q}(t)$, but more generally, $\vec P$ can be selected to provide a physically insightful definition. Differentiating Eq.~\eqref{eq:reconstruct_principle_momentum} with respect to time, and substituting the acceleration from Eq.~\eqref{eq:reconstruct_principle_acceleration} recasts the equation of motion into the form:
\begin{align}
\dot{\vec \pi}(t) = \vec F(\vec q, \dot{\vec q}, t) \text, \label{eq:reconstruct_principle_pi_dot}
\end{align}
where
\begin{equation}
    \vec F \equiv \Dt[\vec P] = \frac{\partial \vec P}{\partial t} + \dot{\vec q} \cdot \frac{\partial \vec P}{\partial \vec q} + \vec U \cdot \frac{\partial \vec P}{\partial \dot{\vec q}}
\end{equation}
is the on-shell force evaluated in the physical limit.

\subsection{Lagrangian}

Consider the linearized Lagrangian in Lie form [Eq.\ \eqref{eq:linearized_action_Lie_form}]. Upon substituting the expressions of $\vec\pi$ and $\Dt[\vec \pi]$ [via Eqs.\ \eqref{eq:reconstruct_principle_momentum} and \eqref{eq:reconstruct_principle_pi_dot}], we get
\begin{equation}
\lagext(\vec q_\pm, \dot{\vec q}_\pm, t) = \dot{\vec q}_- \cdot \vec P(\vec q_+, \dot{\vec q}_+, t) + \vec q_- \cdot \vec F(\vec q_+, \dot{\vec q}_+, t). \label{eq:reconstructed_Lagrangian}
\end{equation}
This corresponds to a kinetic term and a nonconservative force term. The trivial choice $\vec P(\vec q, \dot{\vec q}, t) = \dot{\vec q}$ can reduce the Lagrangian back to form \eqref{eq:reconstructed_Lagrangian_simple_2}. 
The physical Euler-Lagrange equation of Lagrangian \eqref{eq:reconstructed_Lagrangian} reduces to:
\begin{equation}
\big(\ddot{\vec q} - \vec U(\vec q, \dot{\vec q}, t)\big) \cdot \frac{\partial \vec P}{\partial \dot{\vec q}} = 0,
\end{equation}
and since $\partial \vec P / \partial \dot{\vec q}$ is non-singular, the original ODE [Eq.~\eqref{eq:reconstruct_principle_acceleration}] is recovered.

\subsection{Hamiltonian}

Similarly, we may reconstruct a Hamiltonian generator $\hamext$ that reproduces the correct canonical equations in the physical limit. Introducing the doubled phase space $(\vec q_a, \vec \pi_a)$, the linear expansion in minus variables provides a general form:
\begin{equation}
\hamext(\vec q_\pm,\vec \pi_\pm, t) = \vec \pi_{-} \cdot \hamext_{\vec \pi}(\vec q_{+},\vec \pi_{+}, t) + \vec q_{-} \cdot \hamext_{\vec q}(\vec q_{+},\vec \pi_{+},t) \text,
\end{equation}
where at this stage the coefficients $\hamext_{\vec q}$ and $\hamext_{\vec \pi}$ are free. The canonical structure in plus-minus variables enforces the on-shell relationships
\begin{align}
\dot{\vec q}_{+} = \lPB\vec q_{+}, \hamext\rPB = \hamext_{\vec \pi}, \qquad 
\dot{\vec \pi}_{+} = \lPB\vec \pi_{+}, \hamext\rPB = -\hamext_{\vec q} \text.
\end{align}
A comparison with Eqs.~(\ref{eq:reconstruct_principle_velocity}, \ref{eq:reconstruct_principle_pi_dot}) then allows us to identify in the physical limit:
\begin{subequations}
\begin{align}    
\hamext_{\vec \pi}(\vec q, \vec \pi, t) &= \vec V(\vec q, \vec \pi, t) \text, \\ 
\hamext_{\vec q}(\vec q, \vec \pi, t) &= - \vec G(\vec q_{+}, \vec \pi_{+}, t) \text,
\end{align}
\end{subequations}
where we employ a Legendre-like procedure to express $\vec F$ in phase space variables
\begin{equation}
    \vec G(\vec q, \vec \pi, t) = \vec F(\vec q, \vec V(\vec q, \vec \pi, t), t) \text.
\end{equation}
From these identifications we construct the Hamiltonian
\begin{equation}
\hamext(\vec q_\pm,\vec \pi_\pm, t) = \vec \pi_{-} \cdot \vec V(\vec q_+, \vec \pi_+, t) - \vec q_{-} \cdot \vec G(\vec q_{+}, \vec \pi_{+}, t) \text, \label{eq:reconstructed_Hamiltonian}
\end{equation}
which is exactly the Lie form [Eq.\ \eqref{eq:linear_hamext_VG}].
The extended system dynamics defined via $\hamext$ is clearly canonical in the doubled variables. Once more, the special case $\vec P = \dot{\vec q}$ leads to a convenient form:
\begin{equation}
\hamext(\vec q_\pm,\vec \pi_\pm, t) = \vec \pi_{-} \cdot \vec \pi_+ - \vec q_{-} \cdot \vec U(\vec q_{+}, \vec \pi_{+}, t),
\end{equation}
which is analogous to the special Lagrangian constructed above.

As discussed, the reconstructed Hamiltonian and Lagrangian in this section are the Lie forms determined in Sec.\ \ref{sec:linear_act} and \ref{sec:linear_Ham}, and are directly related by a Legendre transformation.
Since higher-order minus terms do not contribute to the dynamics, these reconstructions are exact. The main underlying ambiguity in the procedure lies in the choice of momentum variable [Eq.\ \eqref{eq:reconstruct_principle_momentum}].




\section{Conclusion}

This work has provided a systematic recipe for generalizing Hamiltonian mechanics to nonconservative systems by extending \citeauthor{Galley2013}'s variable doubling technique.
A central aspect of the formulation is embedding the non–conservative dynamics into a higher–dimensional Hamiltonian framework; globally, the phase–space volume is preserved, while local dissipative behaviour emerges on the lower–dimensional physical slice. In this way, the apparent contradiction between dissipation and Hamiltonian mechanics is reconciled.  Although similar embeddings have been contemplated before \cite[see e.g.][]{Smale1970}, our construction is distinctive in that it arises through an ordinary Legendre transform [Eqs. \eqref{eq:nonconservative_Legendre_transform} and \eqref{eq:linear_hamext}] of a well–posed Lagrangian (within an initial value action principle) and retains a direct connection to physical quantities. 


Because the framework automatically assures canonicity, non-conservative terms can be appended to familiar conservative Hamiltonians without requiring adapted analytical machinery [see Eqs.\ \eqref{eq:hamext} and \eqref{eq:hamext_alt}]. This provides a particularly versatile bridge for connecting symplectic and non-symplectic systems.
In particular, we highlight its effectiveness in applications where dissipation enters perturbatively.

Another important strength of our formulation arises from the canonicalisation techniques detailed in Sec.~\ref{sec:inverse-variational-problem}. These methods enable the application of Hamiltonian tools to arbitrary second-order non-symplectic systems via the systematical derivation of double variable Hamiltonians. Similarly, we have provided a recipe for reconstructing the corresponding double variable Lagrangian, which is related by Legendre transform. These capabilities are particularly valuable because they broaden the scope of Hamiltonian methods to previously inaccessible dissipative contexts. 
Nonetheless, as in standard mechanics, the regularity assumptions for reconstructing a Hamiltonian or Lagrangian might not hold globally for certain systems, thereby restricting the domain of applicability.

We have also provided general insights into Galley's principle of nonconservative action. In particular, we have clarified that the physical limit emerges naturally from the doubled variational setup when appropriate initial conditions are imposed---a result that holds provided the Lagrangian or Hamiltonian satisfies the necessary smoothness properties. We recognise that this requirement may fail in certain constrained systems, as illustrated by classical examples such as the pencil-on-a-cusp problem.
We have also clarified the roles of different boundary conditions to the initial value setup [see Eq.~\eqref{eq:boundary_conditions_pmv}].

Finally, we have shown that an essential feature of the doubling of degrees of freedom is the ability to shift the momentum by an arbitrary function of physical variables. This transformation keeps the dynamics invariant but suitably corrects the Hamiltonian so that the symplectic structure is retained [see Eqs.~\eqref{eq:hamext_gauge} and \eqref{eq:hamext_gauge_cons}].
This freedom reveals interesting toy applications even in the case of purely conservative systems. For example, in electromagnetic (EM) contexts---such as in the dynamics of charged particles---the canonical momentum receives an additional term proportional to the vector potential $\vec A$. This term emerges as a shift in the arguments of the Hamiltonian which can be precisely negated by an appropriate canonical gauge transformation. As a consequence, the phase-space dynamics can be directly expressed in terms of the kinetic momentum, potentially simplifying manipulations and results. 
Although the gauge freedom offers flexibility, we also highlight that it introduces an ambiguity in the dissipative components of the conjugate momenta, which demand additional physical insight to resolve in concrete applications. A natural resolution is to maintain the conjugate momentum equal to its conservative definition.

The theoretical groundwork laid here opens several avenues for future investigation. A promising direction is the study of numerical methods---such as symplectic or variational integrators---tailored to dissipative dynamics \cite[see e.g.][]{Tsang2015}. The practical utility of these methods will depend on their stability properties, particularly whether small deviations from the physical slice may lead to divergent or chaotic behaviour. 
In addition, exploring connections with other non–conservative formulations, such as contact geometry, may reveal deeper mathematical relationships and inspire the development of unifying frameworks that capitalise on the strengths of each approach. 

\begin{acknowledgments}

C.A. acknowledges the joint financial support of Centre National d'Études Spatiales (CNES) and École Doctorale Astronomie et Astrophysique d'Ile de France (ED127 AAIF). This work was also supported by PNPS (CNRS/INSU), by INP and IN2P3 co-funded by CNES. We are very grateful to A.\ Albouy, S.\ Bouquillon, and G.\ Faye for the constructive comments and discussions.
\end{acknowledgments}

\newpage

\bibliographystyle{apsrev4-2}
\bibliography{refs}

\begin{thebibliography}{27}%
\makeatletter
\providecommand \@ifxundefined [1]{%
 \@ifx{#1\undefined}
}%
\providecommand \@ifnum [1]{%
 \ifnum #1\expandafter \@firstoftwo
 \else \expandafter \@secondoftwo
 \fi
}%
\providecommand \@ifx [1]{%
 \ifx #1\expandafter \@firstoftwo
 \else \expandafter \@secondoftwo
 \fi
}%
\providecommand \natexlab [1]{#1}%
\providecommand \enquote  [1]{``#1''}%
\providecommand \bibnamefont  [1]{#1}%
\providecommand \bibfnamefont [1]{#1}%
\providecommand \citenamefont [1]{#1}%
\providecommand \href@noop [0]{\@secondoftwo}%
\providecommand \href [0]{\begingroup \@sanitize@url \@href}%
\providecommand \@href[1]{\@@startlink{#1}\@@href}%
\providecommand \@@href[1]{\endgroup#1\@@endlink}%
\providecommand \@sanitize@url [0]{\catcode `\\12\catcode `\$12\catcode
  `\&12\catcode `\#12\catcode `\^12\catcode `\_12\catcode `\%12\relax}%
\providecommand \@@startlink[1]{}%
\providecommand \@@endlink[0]{}%
\providecommand \url  [0]{\begingroup\@sanitize@url \@url }%
\providecommand \@url [1]{\endgroup\@href {#1}{\urlprefix }}%
\providecommand \urlprefix  [0]{URL }%
\providecommand \Eprint [0]{\href }%
\providecommand \doibase [0]{https://doi.org/}%
\providecommand \selectlanguage [0]{\@gobble}%
\providecommand \bibinfo  [0]{\@secondoftwo}%
\providecommand \bibfield  [0]{\@secondoftwo}%
\providecommand \translation [1]{[#1]}%
\providecommand \BibitemOpen [0]{}%
\providecommand \bibitemStop [0]{}%
\providecommand \bibitemNoStop [0]{.\EOS\space}%
\providecommand \EOS [0]{\spacefactor3000\relax}%
\providecommand \BibitemShut  [1]{\csname bibitem#1\endcsname}%
\let\auto@bib@innerbib\@empty
\bibitem [{\citenamefont {Bauer}(1931)}]{Bauer1931}%
  \BibitemOpen
  \bibfield  {author} {\bibinfo {author} {\bibfnamefont {P.~S.}\ \bibnamefont
  {Bauer}},\ }\href {https://doi.org/10.1073/pnas.17.5.311} {\bibfield
  {journal} {\bibinfo  {journal} {Proceedings of the National Academy of
  Sciences}\ }\textbf {\bibinfo {volume} {17}},\ \bibinfo {pages} {311}
  (\bibinfo {year} {1931})}\BibitemShut {NoStop}%
\bibitem [{\citenamefont {Gurtin}(1963)}]{Gurtin1963}%
  \BibitemOpen
  \bibfield  {author} {\bibinfo {author} {\bibfnamefont {M.~E.}\ \bibnamefont
  {Gurtin}},\ }\href {https://doi.org/https://doi.org/10.1007/BF00248489}
  {\emph {\bibinfo {title} {Variational principles for linear
  elastodynamics}}},\ Vol.~\bibinfo {volume} {16}\ (\bibinfo  {publisher}
  {Brown University},\ \bibinfo {year} {1963})\ pp.\ \bibinfo {pages}
  {34--50}\BibitemShut {NoStop}%
\bibitem [{\citenamefont {Gurtin}(1964)}]{Gurtin1964}%
  \BibitemOpen
  \bibfield  {author} {\bibinfo {author} {\bibfnamefont {M.~E.}\ \bibnamefont
  {Gurtin}},\ }\href {https://doi.org/https://doi.org/10.1090/qam/99951}
  {\bibfield  {journal} {\bibinfo  {journal} {Quarterly of Applied
  Mathematics}\ }\textbf {\bibinfo {volume} {22}},\ \bibinfo {pages} {252}
  (\bibinfo {year} {1964})}\BibitemShut {NoStop}%
\bibitem [{\citenamefont {Tonti}(1973)}]{Tonti1973}%
  \BibitemOpen
  \bibfield  {author} {\bibinfo {author} {\bibfnamefont {E.}~\bibnamefont
  {Tonti}},\ }\href {https://doi.org/https://doi.org/10.1007/BF02410725}
  {\bibfield  {journal} {\bibinfo  {journal} {Annali di Matematica pura ed
  applicata}\ }\textbf {\bibinfo {volume} {95}},\ \bibinfo {pages} {331}
  (\bibinfo {year} {1973})}\BibitemShut {NoStop}%
\bibitem [{\citenamefont {Dargush}\ and\ \citenamefont
  {Kim}(2012)}]{Dargush2012}%
  \BibitemOpen
  \bibfield  {author} {\bibinfo {author} {\bibfnamefont {G.~F.}\ \bibnamefont
  {Dargush}}\ and\ \bibinfo {author} {\bibfnamefont {J.}~\bibnamefont {Kim}},\
  }\href {https://doi.org/10.1103/PhysRevE.85.066606} {\bibfield  {journal}
  {\bibinfo  {journal} {Physical Review E—Statistical, Nonlinear, and Soft
  Matter Physics}\ }\textbf {\bibinfo {volume} {85}},\ \bibinfo {pages}
  {066606} (\bibinfo {year} {2012})}\BibitemShut {NoStop}%
\bibitem [{\citenamefont {Dargush}(2012)}]{Dargush2012a}%
  \BibitemOpen
  \bibfield  {author} {\bibinfo {author} {\bibfnamefont {G.~F.}\ \bibnamefont
  {Dargush}},\ }\href {https://doi.org/10.1103/PhysRevE.86.066606} {\bibfield
  {journal} {\bibinfo  {journal} {Phys. Rev. E}\ }\textbf {\bibinfo {volume}
  {86}},\ \bibinfo {pages} {066606} (\bibinfo {year} {2012})}\BibitemShut
  {NoStop}%
\bibitem [{\citenamefont {Bateman}(1931)}]{Bateman1931}%
  \BibitemOpen
  \bibfield  {author} {\bibinfo {author} {\bibfnamefont {H.}~\bibnamefont
  {Bateman}},\ }\href {https://doi.org/10.1103/PhysRev.38.815} {\bibfield
  {journal} {\bibinfo  {journal} {Phys. Rev.}\ }\textbf {\bibinfo {volume}
  {38}},\ \bibinfo {pages} {815} (\bibinfo {year} {1931})}\BibitemShut
  {NoStop}%
\bibitem [{\citenamefont {Schwinger}(1961)}]{Schwinger1961}%
  \BibitemOpen
  \bibfield  {author} {\bibinfo {author} {\bibfnamefont {J.}~\bibnamefont
  {Schwinger}},\ }\href {https://doi.org/10.1063/1.1703727} {\bibfield
  {journal} {\bibinfo  {journal} {Journal of Mathematical Physics}\ }\textbf
  {\bibinfo {volume} {2}},\ \bibinfo {pages} {407} (\bibinfo {year}
  {1961})}\BibitemShut {NoStop}%
\bibitem [{\citenamefont {Keldysh}(1965)}]{Keldysh1964}%
  \BibitemOpen
  \bibfield  {author} {\bibinfo {author} {\bibfnamefont {L.~V.}\ \bibnamefont
  {Keldysh}},\ }\href {https://doi.org/10.1142/9789811279461_0007} {\bibfield
  {journal} {\bibinfo  {journal} {Sov. Phys. JETP}\ }\textbf {\bibinfo {volume}
  {20}},\ \bibinfo {pages} {1018} (\bibinfo {year} {1965})}\BibitemShut
  {NoStop}%
\bibitem [{\citenamefont {Galley}(2013)}]{Galley2013}%
  \BibitemOpen
  \bibfield  {author} {\bibinfo {author} {\bibfnamefont {C.~R.}\ \bibnamefont
  {Galley}},\ }\bibfield  {journal} {\bibinfo  {journal} {Physical Review
  Letters}\ }\textbf {\bibinfo {volume} {110}},\ \href
  {https://doi.org/10.1103/physrevlett.110.174301}
  {10.1103/physrevlett.110.174301} (\bibinfo {year} {2013})\BibitemShut
  {NoStop}%
\bibitem [{\citenamefont {Riewe}(1996)}]{Riewe1996}%
  \BibitemOpen
  \bibfield  {author} {\bibinfo {author} {\bibfnamefont {F.}~\bibnamefont
  {Riewe}},\ }\href {https://doi.org/10.1103/PhysRevE.53.1890} {\bibfield
  {journal} {\bibinfo  {journal} {Phys. Rev. E}\ }\textbf {\bibinfo {volume}
  {53}},\ \bibinfo {pages} {1890} (\bibinfo {year} {1996})}\BibitemShut
  {NoStop}%
\bibitem [{\citenamefont {Riewe}(1997)}]{Riewe1997}%
  \BibitemOpen
  \bibfield  {author} {\bibinfo {author} {\bibfnamefont {F.}~\bibnamefont
  {Riewe}},\ }\href {https://doi.org/10.1103/PhysRevE.55.3581} {\bibfield
  {journal} {\bibinfo  {journal} {Phys. Rev. E}\ }\textbf {\bibinfo {volume}
  {55}},\ \bibinfo {pages} {3581} (\bibinfo {year} {1997})}\BibitemShut
  {NoStop}%
\bibitem [{\citenamefont {{Caldirola}}(1941)}]{Caldirola1941}%
  \BibitemOpen
  \bibfield  {author} {\bibinfo {author} {\bibfnamefont {P.}~\bibnamefont
  {{Caldirola}}},\ }\href {https://doi.org/10.1007/BF02960144} {\bibfield
  {journal} {\bibinfo  {journal} {Il Nuovo Cimento}\ }\textbf {\bibinfo
  {volume} {18}},\ \bibinfo {pages} {393} (\bibinfo {year} {1941})}\BibitemShut
  {NoStop}%
\bibitem [{\citenamefont {Kanai}(1948)}]{Kanai1948}%
  \BibitemOpen
  \bibfield  {author} {\bibinfo {author} {\bibfnamefont {E.}~\bibnamefont
  {Kanai}},\ }\href {https://doi.org/10.1143/ptp/3.4.440} {\bibfield  {journal}
  {\bibinfo  {journal} {Progress of Theoretical Physics}\ }\textbf {\bibinfo
  {volume} {3}},\ \bibinfo {pages} {440} (\bibinfo {year} {1948})}\BibitemShut
  {NoStop}%
\bibitem [{\citenamefont {Calzetta}\ and\ \citenamefont
  {Hu}(2008)}]{Calzetta2008}%
  \BibitemOpen
  \bibfield  {author} {\bibinfo {author} {\bibfnamefont {E.~A.}\ \bibnamefont
  {Calzetta}}\ and\ \bibinfo {author} {\bibfnamefont {B.-L.~B.}\ \bibnamefont
  {Hu}},\ }\href@noop {} {\emph {\bibinfo {title} {Nonequilibrium Quantum Field
  Theory}}},\ Cambridge Monographs on Mathematical Physics\ (\bibinfo
  {publisher} {Cambridge University Press},\ \bibinfo {year}
  {2008})\BibitemShut {NoStop}%
\bibitem [{\citenamefont {Rayleigh}\ and\ \citenamefont
  {Baron}(1896)}]{Rayleigh1896}%
  \BibitemOpen
  \bibfield  {author} {\bibinfo {author} {\bibfnamefont {J.~S.~B.}\
  \bibnamefont {Rayleigh}}\ and\ \bibinfo {author} {\bibfnamefont
  {S.}~\bibnamefont {Baron}},\ }\href
  {https://gallica.bnf.fr/ark:/12148/bpt6k95131k} {\emph {\bibinfo {title} {The
  theory of sound, vol. 2, revised and enlarged macmillan}}},\ Vol.\ \bibinfo
  {volume} {132}\ (\bibinfo  {publisher} {London: Macmillan and Co.},\ \bibinfo
  {year} {1896})\BibitemShut {NoStop}%
\bibitem [{\citenamefont {Virga}(2015)}]{Virga2015}%
  \BibitemOpen
  \bibfield  {author} {\bibinfo {author} {\bibfnamefont {E.~G.}\ \bibnamefont
  {Virga}},\ }\href {https://doi.org/10.1103/PhysRevE.91.013203} {\bibfield
  {journal} {\bibinfo  {journal} {Phys. Rev. E}\ }\textbf {\bibinfo {volume}
  {91}},\ \bibinfo {pages} {013203} (\bibinfo {year} {2015})}\BibitemShut
  {NoStop}%
\bibitem [{\citenamefont {Damour}\ \emph {et~al.}(2016)\citenamefont {Damour},
  \citenamefont {Jaranowski},\ and\ \citenamefont {Sch\"afer}}]{Damour2016}%
  \BibitemOpen
  \bibfield  {author} {\bibinfo {author} {\bibfnamefont {T.}~\bibnamefont
  {Damour}}, \bibinfo {author} {\bibfnamefont {P.}~\bibnamefont {Jaranowski}},\
  and\ \bibinfo {author} {\bibfnamefont {G.}~\bibnamefont {Sch\"afer}},\ }\href
  {https://doi.org/10.1103/PhysRevD.93.084014} {\bibfield  {journal} {\bibinfo
  {journal} {Phys. Rev. D}\ }\textbf {\bibinfo {volume} {93}},\ \bibinfo
  {pages} {084014} (\bibinfo {year} {2016})}\BibitemShut {NoStop}%
\bibitem [{\citenamefont {Bravetti}\ \emph {et~al.}(2017)\citenamefont
  {Bravetti}, \citenamefont {Cruz},\ and\ \citenamefont
  {Tapias}}]{Bravetti2017}%
  \BibitemOpen
  \bibfield  {author} {\bibinfo {author} {\bibfnamefont {A.}~\bibnamefont
  {Bravetti}}, \bibinfo {author} {\bibfnamefont {H.}~\bibnamefont {Cruz}},\
  and\ \bibinfo {author} {\bibfnamefont {D.}~\bibnamefont {Tapias}},\ }\href
  {https://doi.org/10.1016/j.aop.2016.11.003} {\bibfield  {journal} {\bibinfo
  {journal} {Annals of Physics}\ }\textbf {\bibinfo {volume} {376}},\ \bibinfo
  {pages} {17} (\bibinfo {year} {2017})}\BibitemShut {NoStop}%
\bibitem [{\citenamefont {Dekker}(1981)}]{Dekker1981}%
  \BibitemOpen
  \bibfield  {author} {\bibinfo {author} {\bibfnamefont {H.}~\bibnamefont
  {Dekker}},\ }\href
  {https://doi.org/https://doi.org/10.1016/0370-1573(81)90033-8} {\bibfield
  {journal} {\bibinfo  {journal} {Phys. Rep.}\ }\textbf {\bibinfo {volume}
  {80}},\ \bibinfo {pages} {1} (\bibinfo {year} {1981})}\BibitemShut {NoStop}%
\bibitem [{\citenamefont {Kosyakov}(2007)}]{Kosyakov2007}%
  \BibitemOpen
  \bibfield  {author} {\bibinfo {author} {\bibfnamefont {B.}~\bibnamefont
  {Kosyakov}},\ }\href {https://doi.org/10.1007/978-3-540-40934-2} {\emph
  {\bibinfo {title} {Introduction to the classical theory of particles and
  fields}}}\ (\bibinfo  {publisher} {Springer Science \& Business Media},\
  \bibinfo {year} {2007})\BibitemShut {NoStop}%
\bibitem [{\citenamefont {Morse}\ and\ \citenamefont
  {Feshbach}(1953)}]{Morse1953}%
  \BibitemOpen
  \bibfield  {author} {\bibinfo {author} {\bibfnamefont {P.~M.}\ \bibnamefont
  {Morse}}\ and\ \bibinfo {author} {\bibfnamefont {H.}~\bibnamefont
  {Feshbach}},\ }\href@noop {} {\emph {\bibinfo {title} {Methods of Theoretical
  Physics. Vol. 1-2}}}\ (\bibinfo  {publisher} {McGraw-Hill},\ \bibinfo {year}
  {1953})\BibitemShut {NoStop}%
\bibitem [{\citenamefont {Um}\ \emph {et~al.}(2002)\citenamefont {Um},
  \citenamefont {Yeon},\ and\ \citenamefont {George}}]{Um2002}%
  \BibitemOpen
  \bibfield  {author} {\bibinfo {author} {\bibfnamefont {C.-I.}\ \bibnamefont
  {Um}}, \bibinfo {author} {\bibfnamefont {K.-H.}\ \bibnamefont {Yeon}},\ and\
  \bibinfo {author} {\bibfnamefont {T.~F.}\ \bibnamefont {George}},\ }\href
  {https://doi.org/https://doi.org/10.1016/S0370-1573(01)00077-1} {\bibfield
  {journal} {\bibinfo  {journal} {Phys. Rep.}\ }\textbf {\bibinfo {volume}
  {362}},\ \bibinfo {pages} {63} (\bibinfo {year} {2002})}\BibitemShut
  {NoStop}%
\bibitem [{\citenamefont {Galley}\ and\ \citenamefont
  {Tiglio}(2009)}]{Galley2009}%
  \BibitemOpen
  \bibfield  {author} {\bibinfo {author} {\bibfnamefont {C.~R.}\ \bibnamefont
  {Galley}}\ and\ \bibinfo {author} {\bibfnamefont {M.}~\bibnamefont
  {Tiglio}},\ }\href {https://doi.org/10.1103/PhysRevD.79.124027} {\bibfield
  {journal} {\bibinfo  {journal} {Phys. Rev. D}\ }\textbf {\bibinfo {volume}
  {79}},\ \bibinfo {pages} {124027} (\bibinfo {year} {2009})}\BibitemShut
  {NoStop}%
\bibitem [{\citenamefont {Galley}\ \emph {et~al.}(2014)\citenamefont {Galley},
  \citenamefont {Tsang},\ and\ \citenamefont {Stein}}]{Galley2014}%
  \BibitemOpen
  \bibfield  {author} {\bibinfo {author} {\bibfnamefont {C.~R.}\ \bibnamefont
  {Galley}}, \bibinfo {author} {\bibfnamefont {D.}~\bibnamefont {Tsang}},\ and\
  \bibinfo {author} {\bibfnamefont {L.~C.}\ \bibnamefont {Stein}},\ }\href
  {https://doi.org/10.48550/arxiv.1412.3082} {\bibinfo {title} {The principle
  of stationary nonconservative action for classical mechanics and field
  theories}} (\bibinfo {year} {2014}),\ \Eprint
  {https://arxiv.org/abs/1412.3082} {arXiv:1412.3082 [math-ph]} \BibitemShut
  {NoStop}%
\bibitem [{\citenamefont {{Smale}}(1970)}]{Smale1970}%
  \BibitemOpen
  \bibfield  {author} {\bibinfo {author} {\bibfnamefont {S.}~\bibnamefont
  {{Smale}}},\ }\href {https://doi.org/10.1007/BF01389805} {\bibfield
  {journal} {\bibinfo  {journal} {Inventiones Mathematicae}\ }\textbf {\bibinfo
  {volume} {11}},\ \bibinfo {pages} {45} (\bibinfo {year} {1970})}\BibitemShut
  {NoStop}%
\bibitem [{\citenamefont {Tsang}\ \emph {et~al.}(2015)\citenamefont {Tsang},
  \citenamefont {Galley}, \citenamefont {Stein},\ and\ \citenamefont
  {Turner}}]{Tsang2015}%
  \BibitemOpen
  \bibfield  {author} {\bibinfo {author} {\bibfnamefont {D.}~\bibnamefont
  {Tsang}}, \bibinfo {author} {\bibfnamefont {C.~R.}\ \bibnamefont {Galley}},
  \bibinfo {author} {\bibfnamefont {L.~C.}\ \bibnamefont {Stein}},\ and\
  \bibinfo {author} {\bibfnamefont {A.}~\bibnamefont {Turner}},\ }\href
  {https://doi.org/10.1088/2041-8205/809/1/L9} {\bibfield  {journal} {\bibinfo
  {journal} {ApJL}\ }\textbf {\bibinfo {volume} {809}},\ \bibinfo {pages} {L9}
  (\bibinfo {year} {2015})}\BibitemShut {NoStop}%
\end{thebibliography}%

\newpage
\appendix
\section{Helmholtz decomposition}
\label{sec:Helmholtz}
Consider a smooth function, antisymmetric under label-exchange, decomposed into the form 
\begin{equation} f(\vec x_\pm) = \vec x_- \cdot \vec{F}(\vec x_+) + \mathcal O(-)^3, \end{equation}
assuming $\vec{F}$ a square-integrable vector field. In analogy with the familiar Helmholtz decomposition in $\reals^3$, $\vec{F}$ can be decomposed into the sum of a gradient and a divergence-free term:
\begin{equation} 
\vec F(\vec x_+) = \grad \phi(\vec x_+) + \vec{\mathcal N}(\vec x_+), \quad 
\text{with} \quad \grad \cdot \vec{\mathcal N} = 0. 
\label{eq:helmholtz_decomp}
\end{equation}
We refer to these as the conservative and nonconservative components respectively. Observe that $\grad \phi$ is a conservative field and admits reconstruction via a path-independent line integral
\begin{equation}
    \int_{\pathB{\vec x}}^{\pathA{\vec x}} \grad \phi(\vec x) \cdot \dd \vec x = \phi(\pathA{\vec x}) - \phi(\pathB{\vec x}) \text,
\end{equation}
so that in the original function $f$ it decouples the two variables $\pathA{\vec x}$ and $\pathB{\vec x}$:
\begin{equation}
    f(\vec x_\pm) = \phi(\pathA{\vec x}) - \phi(\pathB{\vec x}) + \vec x_- \cdot \vec{\mathcal N}(\vec x_+) + \mathcal O(-)^3 \text.
\end{equation}
Hence, the dynamics of $\phi$ would be typically expressible from the standard conservative framework. Meanwhile, the solenoidal term $\vec{\mathcal N}$, which is not similarly separable, must in general describe nonconservative dynamics.
In $\mathbb{R}^3$, $\vec{\mathcal N}$ can also be written as the curl of some vector potential $\vec A$, that is, $\vec{\mathcal N} = \grad \times \vec A$. More generally in $\reals^{2N}$, one must adopt the expression
$\vec{\mathcal N} = \star \dd \vec A$, with $\star$ the Hodge dual operator. In component notation this becomes:
\begin{equation}
    \mathcal N_i = \frac{1}{(2N-2)!}\epsilon_{i j k_1 \ldots k_{2N-2}} \partial^{j} A^{k_1 \ldots k_{2N-2}} \text.
\end{equation}

To perform the Helmholtz decomposition of $\vec F$, one takes the divergence of Eq.~\eqref{eq:helmholtz_decomp}, so as to recover the Poisson equation for $\phi$
\begin{equation}
    \laplac \phi = \grad \cdot \vec F \text.
\end{equation}
The solution to this equation is given by convolution with the Green’s function $G$ for the Laplacian in $\reals^{2N}$:
\begin{equation}
    \phi(\vec{x}) = \int_{\mathbb{R}^{2N}} G(\vec{x}-\vec{x}\,')\, \grad \cdot \vec{F}(\vec{x}\,') \, \dd^{2N}\vec{x}\,'.
\end{equation}

The appropriate Green’s function must be selected to respect the domain’s boundaries. Under the assumption that the scalar and vector potentials decay at infinity, the decomposition is unique.
For $2N \geq 3$, the fundamental solution is
\begin{equation}
    G(\vec x) = \frac{1}{(2N-2) S_{2N}} \frac{1}{x^{2N-2}} \text,
\end{equation}
where $S_{2N}$ is the surface area of the unit sphere in $\reals^{2N}$, namely
\begin{equation}
    S_{2N} = \frac{2 \pi^N}{\Gamma(N)} \text.
\end{equation}
Note that for $2N=2$ the form of $G$ is logarithmic.

Once $\phi$ is known, the nonconservative part of $\vec F$ is given simply by subtracting the gradient component:
\begin{equation}
    \vec{\mathcal N} = \vec F - \grad \phi \text,
\end{equation}
which is divergence-free by construction. 
In domains with general boundary conditions (for example, on a bounded domain with Dirichlet or Neumann conditions), the uniqueness of the decomposition holds only up to the addition of harmonic functions that satisfy the boundary conditions. In particular, if the decay conditions at infinity are imposed, no such (nonzero) harmonic functions exist and the decomposition is unique. We recall that a harmonic function $\lambda$ is a scalar field satisfying $\laplac \lambda = 0$. Indeed, by shifting the scalar potential by $\phi \mapsto \phi + \lambda$, and accordingly the nonconservative field by $\vec{\mathcal N} \mapsto \vec{\mathcal N} - \grad \lambda$, another solution is generated. The harmonic condition on $\lambda$ ensures the preservation of the divergence-free property of $\vec{\mathcal N}$.

When the decomposition is expressed in terms of the vector potential $\vec A$, an additional freedom emerges, which is the choice of gauge $\vec A \mapsto \vec A + \grad \varphi$. Clearly, this gauge choice does not affect the split between the conservative and nonconservative parts $\grad \phi$ and $\vec{\mathcal N}$, unlike the addition of harmonic terms $\lambda$.

In the double variable framework, the Helmholtz decomposition of the Lagrangian $\lagext$ with $\vec x = (\vec q, \dot{\vec q})$ implies a splitting between conservative and nonconservative contributions into $\lagstd$ and $\mathcal K$, which is not unique unless additional boundary criteria are supposed. Nevertheless, a clear consequence is that for $\mathcal K$ to describe nonconservative phenomena, then a necessary condition is that every Helmholtz decomposition must have $\mathcal N \neq 0$; in other words, $\mathcal K$ must not be expressible as the gradient of a potential with respect to $\vec x$. Similar considerations can be applied to the extended Hamiltonian presented in Sec.~\ref{sec:nonconservative-Hamiltonian-formulation}.

\end{document}